\documentclass[12pt]{article}

\usepackage[super]{natbib}
\usepackage{graphicx}
\usepackage{revsymb}	
\usepackage{amsmath}	
\usepackage{txfonts}
\usepackage{amssymb}
\usepackage{geometry} 
\usepackage[parfill]{parskip} 
\usepackage{slantsc}
\usepackage{array}
\usepackage{url}
\usepackage{amsfonts}
\usepackage[colorlinks=true,linkcolor=black, urlcolor=black, citecolor=black]{hyperref}
\usepackage{hyperref} 
\usepackage{sidecap}
\usepackage{xcolor}
\usepackage{nicefrac}
\usepackage{epsfig}
\colorlet{rouge}{red!70!darkgray}
\usepackage{times}
\usepackage{fancyhdr}

\usepackage{float}
\usepackage{latexsym}
\usepackage{setspace}
\usepackage{comment}
\usepackage{amssymb}
\usepackage{gensymb}
\usepackage{fancyhdr}
\usepackage{nicefrac}
\usepackage{color}
\usepackage[normalem]{ulem}

\newcommand*\aap{A\&A}

\newcommand*\apj{ApJ}
\newcommand*\apjl{ApJ}

\newcommand*\apjs{ApJS}

\newcommand*\apss{Ap\&SS}
\newcommand*\araa{ARA\&A}

\newcommand*\mnras{MNRAS}

\newcommand*\nat{Nature}

\newcommand*\pasj{PASJ}

\newcommand*\prl{Phys.~Rev.~Lett.}

\newcommand*\solphys{Sol.~Phys.}

\newcommand*\ssr{Space~Sci.~Rev.}

\title{Helioseismic inference of the solar radiative opacity} 

\author
{Ga\"el Buldgen$^{1,2\ast}$, Jean-Christophe Pain$^{3,4}$,  Philippe Coss\'e$^{3,4}$, Christophe Blancard$^{3,4}$,\\ 
Franck Gilleron$^{3,4}$, Anil Pradhan$^{5}$, Christopher J. Fontes$^{6}$, James Colgan$^{6}$, \\ Arlette Noels$^{2}$, J\o rgen Christensen-Dalsgaard$^{7}$, Morgan Deal$^{8,9}$,\\ Sergey V. Ayukov$^{10}$,  Vladimir A. Baturin$^{10}$,\\ Anna V. Oreshina$^{10}$,  Richard Scuflaire$^{1}$,\\ Charly Pin\c{c}on$^{11}$, Yveline Lebreton$^{12,13}$,\\ Thierry Corbard$^{14}$, Patrick Eggenberger$^{2}$, Peter Hakel$^{6}$, \\ David P. Kilcrease$^{6}$ \\
\small{$^{1}$Institut d'Astrophysique et G\'eophysique de l'Universit\'e de Li\`ege, All\'ee du 6 ao\^ut 17,}\\
\small{4000 Li\`ege, Belgium}\\
\small{$^{2}$D\'epartement d'Astronomie, Universit\'e de Gen\`eve, Chemin Pegasi 51, CH-1290 Versoix, Switzerland}\\
\small{$^{3}$CEA, DAM, DIF, F-91297 Arpajon, France}\\
\small{$^{4}$Universit\'e Paris-Saclay, CEA, LMCE, F-91680 Bruy\`eres le Ch\^atel, France}\\
\small{$^{5}$Ohio State University, Dept. Astronomy, Columbus, OH, 43210, USA}\\
\small{$^{6}$Los Alamos National Laboratory, Los Alamos, NM, 87545, USA}\\
\small{$^{7}$Stellar Astrophysics Centre and Department of Physics and Astronomy, Aarhus University,}\\
\small{8000, Aarhus C, Denmark}\\
\small{$^{8}$LUPM, Universit\'e de Montpellier, CNRS, Place Eug\`ene Bataillon, 34095 Montpellier, France}\\
\small{$^{9}$Instituto de Astrof\'isica e Ci\^encias do Espa\c{c}o, Universidade do Porto, CAUP, Rua das Estrelas,}\\
\small{PT4150-762 Porto, Portugal}\\
\small{$^{10}$Sternberg Astronomical Institute, Lomonosov Moscow State University, 119234,}\\
\small{Moscow, Russia}\\
\small{$^{11}$Universit\'e Paris-Saclay, Institut d\' Astrophysique Spatiale, UMR 8617, CNRS, B\^atiment 121, 91405,}\\
\small{ Orsay Cedex, France}\\
\small{$^{12}$LESIA, Observatoire de Paris, Universit\'e PSL, CNRS, Sorbonne Universit\'e, Universit\'e Paris Cit\'e,}\\ 
\small{5 place Jules Janssen, 92195 Meudon, France}\\
\small{$^{13}$Univ Rennes, CNRS, IPR (Institut de Physique de Rennes) – UMR 6251, 35000 Rennes, France}\\
\small{$^{14}$Universit\'e C\^ote d'Azur, Observatoire de la C\^ote d'Azur, CNRS,}\\
\small{Laboratoire Lagrange, France}\\
\small{$^\ast$To whom correspondence should be addressed; E-mail:  g.buldgen@uliege.be.}
}

\date{}

\begin{document} 

\newcommand{\bin}[2]{\left(\begin{array}{c}\!#1\!\\\!#2\!\end{array}\right)}
\newcommand{\threejm}[6]{\left(\begin{array}{ccc}#1 & #2 & #3 \\ #4 & #5 & #6 \end{array}\right)}


\baselineskip24pt


\maketitle 
\newpage

{\bf 
The Sun is the most studied of all stars, and thus constitutes a benchmark for stellar models. However, our vision of the Sun is still incomplete, as illustrated by the current debate on its chemical composition. The problem reaches far beyond chemical abundances and is intimately linked to microscopic and macroscopic physical ingredients of solar models such as radiative opacity, for which experimental results have been recently measured that still await theoretical explanations. We present opacity profiles derived from helioseismic inferences and compare them with detailed theoretical computations of individual element contributions using three different opacity computation codes, in a complementary way to experimental results. We find that our seismic opacity is about 10$\%$ higher than theoretical values used in current solar models around 2 million degrees, but lower by 35$\%$ than some recent available theoretical values. Using the Sun as a laboratory of fundamental physics, we show that quantitative comparisons between various opacity tables are required to understand the origin of the discrepancies between reported helioseismic, theoretical and experimental opacity values.
}

\newpage
\section*{Introduction}

The Sun is the most studied star in the Universe. For decades, it has been observed by numerous ground-based and space-based instruments, establishing it as a benchmark in stellar physics. With the advent of helioseismology, the study of solar oscillations, we gained a direct access to its internal structure (see e.g. Christensen-Dalsgaard 2021 \cite{JCD2021} and references therein) and could use the Sun as a laboratory of fundamental physics. The field achieved many breakthroughs: the location of the base of the solar convective envelope ($r_{cz}=0.713\pm0.001\rm{R_{Sun}}$ (hereafter BCZ), \cite{JCD91Conv, Basu97BCZ}), the determination of its internal rotation (e.g. Couvidat et al. 2003\cite{Couvidat2003}, Howe et al. 2009\cite{Howe2009}) and sound speed profiles (e.g. Christensen-Dalsgaard et al. 1985, 1989\cite{JCD1985,JCD1989}, Antia $\&$ Basu 1994\cite{Antia94Nonlin}) as well as the helium mass fraction in the solar envelope ($Y_{cz}=0.2485\pm0.0035$, \cite{BasuYSun,Vorontsov13}), inaccessible to surface spectroscopy. The excellent agreement between standard solar models (SSM) and helioseismic constraints (e.g. Christensen-Dalsgaard et al. 1996\cite{MODELS}) also impacted the resolution of the ``solar neutrino problem''. 

However, the beginning of the $21^{st}$ century brought a $25\%$ downward revision of the solar carbon, nitrogen and oxygen surface abundances \cite{Allende2001,Allende2003,Asplund2004,Asplund2006,AGSS09} with respect to Grevesse $\&$ Sauval 1998\cite{GS98} and $40\%$ with respect to the widely acclaimed solar Model S\cite{MODELS}. Using the updated surface abundance values, this previous agreement was significantly worsened, $r_{cz}$ and $Y_{cz}$ did not fit observations anymore and large discrepancies appeared in sound speed. This defined the ``solar modeling problem'', that is actually not limited to the solar case. In stellar modeling, solar abundances define the ``metallicity
scale'' that relates stellar abundances to solar ones, so that any change of the solar reference impacts stellar astrophysics globally. The solar problem is still extensively studied today (see e.g. Serenelli et al. 2009\cite{Serenelli2009}, Song et al. 2018\cite{Song2018} and Zhang et al. 2019\cite{Zhang2019} and references therein, and Basu 2016\cite{Basu2016} for a review), as it impacts the ingredients entering stellar evolution computations such as, amongst others, the transport of chemical elements, the equation of state or radiative opacities \cite{Guzik2005,Ayukov2010}. The latter were quickly identified as a potential source of the disagreements \cite{Montalban2006,JCD2009,Ayukov2017}, either due to abundance revisions \cite{Antia2005,Zaatri2007} or inaccuracies in opacity computations. A revision of the solar neon abundance was recently determined \cite{Young, Asplund2021}, but is insufficient in reconciling the low metallicity standard models with helioseismic constraints. Magg et al. (2022)\cite{Magg2022} claimed to solve the problem by restoring the agreement between solar models and helioseismic data achieved in the 1990s with revised abundances derived from averaged 3D atmosphere models. However, their solar models neglect rotation and light element depletion \cite{Eggenberger2022} as well as the latest opacity tables published in 2015 and 2016 \cite{Mondet,Colgan}. The oxygen revision they propose has also been recently questioned \cite{Nahar2021} due to issues in the photoionization data used, which was later claimed to have no impact on the oxygen abundance inference\cite{Bautista2022}. Recent helioseismic determinations of the solar heavy-element content favour a lower value \cite{Vorontsov13, Buldgen2017,B2023Z}. Buldgen et al. (2023)\cite{Buldgen2023} showed that the agreement found in Magg et al. (2022)\cite{Magg2022} was a direct result of the standard model recipe and did not alleviate the need for further revisions of fundamental ingredients. Overall, the debate has unveiled a more complex, multi-faceted picture of the current problem with solar models that impacts multiple aspects beyond abundance determinations and thus requires innovative approaches to tackle it and fully exploit helioseismic data. The first measurement of iron opacity in almost solar conditions at Sandia national laboratories \cite{Bailey} showed discrepancies between $30\%$ and $400\%$ with theoretical results in the wavelength range between $7.0$ and $12.7$ $\rm{\AA}$. Further measurements were carried out for nickel and chromium, showing also significant, although more modest, discrepancies \cite{Nagayama2019} while more recent experiments for iron exclude the higher range of values from the Sandia measurements \cite{Hoarty2024}. Early inversions\cite{JCD1985} also hinted at possible opacity revisions in solar models of the time and thus motivate our approach to determine the solar opacity directly from helioseismic data. The current state of affairs is somewhat reminiscent of the 1980s, when various authors\cite{Simon1982,Andreasen1988} pleaded to improve theoretical opacity computations for Cepheids and in line with the conclusions of detailed seismic analyses of massive stars\cite{Moravveji2016b,DD2017}. 

The solar modelling problem is linked to both microphysical and macrophysical aspects of standard solar models (See Christensen-Dalsgaard 2021\cite{JCD2021} for a definition of the standard solar model framework) and thus to how we depict the processes acting in the Sun. Standard models entirely neglect rotation, thus failing to reproduce the observed depletion of lithium in the Sun. Revising the ``recipe'' for solar models implies a revision of the ingredients used for other stars in the Universe as well, particularly impacting the masses, radii and ages inferred from the seismic modelling of low-mass stars. In massive stars, opacity modifications significantly alter the oscillation properties \cite{ZdravkovOpac,SalmonOpac,Aerts2021}. Previous work\cite{Moravveji2016,DD2017} has shown that modifying opacity in the direction of the experiments would improve the agreement with asteroseismic observations.

Recently, Eggenberger et al. (2022)\cite{Eggenberger2022} showed that reproducing the solar lithium abundance also brought the $\rm{Y_{cz}}$ value of low metallicity models in agreement with helioseismic measurements, linking it to the angular momentum transport mechanism allowing to reproduce the solar rotation profile. They showed that the link between lithium and helium was robust with respect to the form and physical origin of the mixing at the BCZ. In this work, we combine these results with the approach of Buldgen et al. (2020)\cite{Buldgen2020} to provide a seismic measurement of the mean Rosseland opacity independently from any theoretical opacity table. Our approach is weakly impacted by the equation of state and nuclear reaction rates used. We are thus able to analyze in details the output of theoretical opacity computations. Technical details on our procedure can be found in the Methods section.
\section*{Results}
\subsection*{Reconstruction of the solar seismic structure and seismic opacity determinations}

Our solar models are computed using the Li\`ege stellar evolution code with the abundances of Asplund et al. (2009)\cite{AGSS09} (hereafter AGSS09); some, like model 8, include the recent neon abundance revision\cite{Young,Asplund2021} (hereafter AAG21) and one, model 7, uses abundances of the 1990's from Grevesse $\%$ Noels (1993) \cite{GrevNoels} (hereafter GN93) whereas model 10 uses the recent abundances\cite{Magg2022} from averaged 3D atmosphere models (hereafter MB22). The properties of the set of calibrated evolutionary models are listed in Table 1, additional information is provided in the Methods section, subsection Solar evolutionary models and macroscopic transport of chemical elements. First, we focus on Models 1 and 2 for the detailed analysis whereas the other models are used in the Methods section to determine whether the observed trends in the seismic opacity profile remain observable for other sets of physical ingredients. All models include microscopic diffusion without the effects of radiative accelerations, as in Buldgen et al. (2019)\cite{Buldgen2019}, as these have been shown \cite{Turcotte1998} to only have a limited impact in the solar case, which would even further be reduced by the effects of macroscopic mixing. Here, Model 7 is essentially a SSM from the 1990's, while we always include the latest physical prescription for macroscopic transport at the BCZ from hydrodynamic and magneto-hydrodynamic instabilities \cite{Eggenberger2022} in other models (using an asymptotic form described in the Methods). We also include adiabatic overshoot at the BCZ so that $r_{cz}$, the location of the base of convective envelope, is located at the helioseismic value. These evolutionary models serve as initial conditions for the procedure of Buldgen et al. (2020)\cite{Buldgen2020}, based on the iterative inversion of the Ledoux discriminant, defined as
\begin{equation}
A = 1/\Gamma_{1} \left(d \ln P/d \ln r\right)-(d \ln \rho/d \ln r) = (\rm{r} \delta/H_{P})\left( \nabla_{\rm{ad}}- \nabla  + (\phi/\delta)\nabla_{\mu} \right)= A_{T}+A_{\mu}, \label{eq:LedouxDisc}
\end{equation}
with $\rho$ the density, $P$ the pressure, $\Gamma_{1}=(\partial \ln P/\partial \ln \rho)\vert_{S}$ the first adiabatic exponent and $S$ the entropy, $r$ the radial position, $\delta=-(\partial \ln \rho/\partial \ln T)$, $H_{P}=-(\rm{d}r/\rm{d} \ln P)$ the pressure scale height, $\nabla_{\rm{ad}}$ the adiabatic temperature gradient, $\nabla=(\rm{d} \ln T/\rm{d} \ln P)$ the temperature gradient, $\phi=(\partial \ln \rho/\partial \ln \mu$) and $\nabla_{\mu}=(\rm{d} \ln \mu/\rm{d} \ln P)$ the mean molecular weight gradient. We separate $A$ in its chemical and thermal components, $A_{\mu}$ and $A_{T}$. 

\begin{table*}[t]
\caption{Parameters of the reference solar evolutionary models.}
\label{tabAddRefModels}
  \centering%
 \resizebox{\linewidth}{!}{%
 \begin{tabular}{|r | c | c | c | c | c | c | c | c |}
\hline 
\textbf{Name}&\textbf{$\left(r_{\rm{BCZ}}/R_{\odot}\right)$}&\textbf{$\mathit{Z}_{\rm{CZ}}/\mathit{X}_{\rm{CZ}}$}&\textbf{$\mathit{Y}_{\rm{CZ}}$}& \textbf{$D_{X,i}$} &\textbf{EOS}&\textbf{Opacity}&\textbf{Relative Abundances} & \textbf{Nuclear reactions}\\ \hline
Model 1&$0.7133$& $0.0181$ & $0.2410$ & $D_{X,1}$ & FreeEOS & OPAS & AGSS09 & Adelberger\\
\hline
Model 2&$0.7133$& $0.0186$ & $0.2486$ & $D_{X,1}$ & SAHA-S & OPAL & AAG21 & Adelberger\\ 
\hline
Model 3&$0.7133$& $0.0181$ & $0.2441$ & $D_{X,1}$ & FreeEOS & OP & AGSS09 & Adelberger\\
\hline
Model 4&$0.7133$& $0.0186$ & $0.2470$ & $D_{X,1}$ & FreeEOS & OPAL & AAG21 & NACRE \\ 
\hline
Model 5&$0.7133$& $0.0181$ & $0.2385$ & $D_{X,1}$ & FreeEOS & OPLIB & AGSS09 & Adelberger\\
\hline
Model 6&$0.7133$& $0.0186$ & $0.2470$ & $D_{X,1}$& FreeEOS & OPAL & AAG21 & Adelberger\\
\hline
Model 7&$0.7133$& $0.0244$ & $0.2457$ & N/A & FreeEOS & OPAL & GN93 & Adelberger\\ 
\hline
Model 8&$0.7133$& $0.0186$ & $0.2467$ & $D_{X,2}$ & FreeEOS & OPAL & AAG21 & Adelberger\\ 
\hline
Model 9&$0.7133$& $0.0186$ & $0.2479$ & $D_{X,1}$ (low) & SAHA-S & OPAL & AAG21 & Adelberger\\ 
\hline
Model 10&$0.7133$& $0.0225$ & $0.2516$ & $D_{X,1}$ & FreeEOS & OP & MB22 & Adelberger\\ 
\hline
\end{tabular}
}
\end{table*}

Figure 1 shows an example of iterative reconstruction using Model 2 as a reference (see Table 1). After convergence, an excellent agreement is reached for all helioseismic constraints whatever the reference model \cite{Buldgen2020} and the resulting structure is therefore called ``seismic model''. The Ledoux discriminant profile resulting from the iterative reconstruction is independent of the initial conditions and provides a direct model-independent measurement of temperature gradients in the Sun, and thus for a given composition, a direct access to measuring the opacity using seismic data. We thus build seismic models using the most recent solar abundances from complete spectra analyses, reproducing the helioseismic structure of the solar radiative zone, the value of $\rm{r_{cz}}$,$\rm{Y_{cz}}$, and lithium, thanks to a formalism reproducing the solar rotation profile in the radiative zone.

\begin{figure*}
	\centering
		\includegraphics[width=15.5cm]{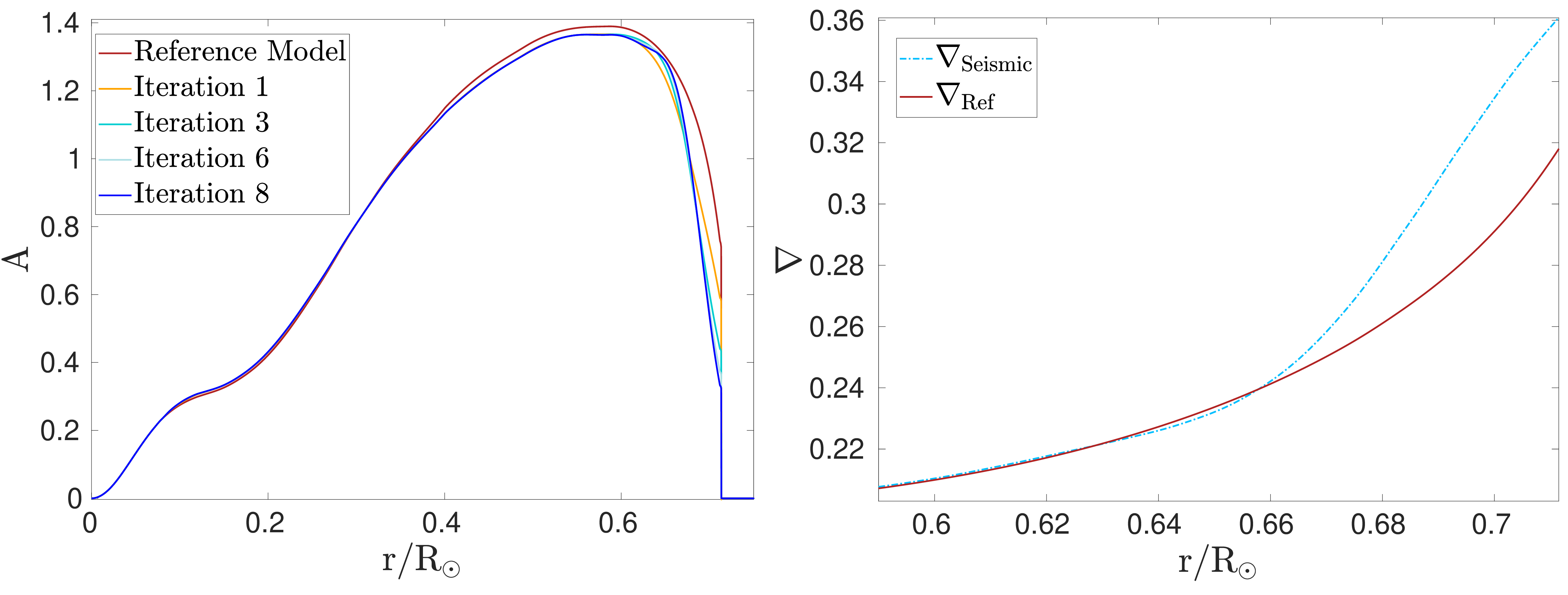}
	\caption{\textbf{Reconstruction procedure of the seismic solar models.} \textit{Left panel:} Iterations on the Ledoux discriminant profile from the seismic reconstruction procedure of \cite{Buldgen2020}. \textit{Right panel:} Zoom on the temperature gradients in the evolutionary model and the seismic model (iteration 8 in the left panel) at the BCZ. Source Data provided with this paper.}
		\label{FigAIteration}
\end{figure*} 

Due to macroscopic turbulence, all effects of microscopic diffusion at the BCZ are smeared (as seen e.g. Brun et al. 2002\cite{Brun02}, Christensen-Dalsgaard et al. 2018\cite{JCD18} for other types of mixing). This is illustrated in Figure 2 for evolutionary models 7 and 9 (see Table 1). Model 7 is a standard solar model. Consequently, it shows deviations at the BCZ between the thermal contribution in dashed orange and the total value in solid red. Model 9, including macroscopic transport, shows a thermal contribution of the Ledoux discriminant ($A_{T}$) almost equal to the total value. The chemical composition profile close to the BCZ is fixed by reproducing simultaneously the surface lithium depletion and $Y_{cz}$ with macroscopic turbulence, as in Model 9.  Thus, we directly measure $A_{T}$ with the inversion. The opacity is then determined from the radiative transfer equation. Thermal equilibrium is ensured by slightly altering the core so that the proper energy amount is generated by the nuclear fusion reactions and the solar luminosity reproduced.

\begin{figure*}
	\centering
		\includegraphics[width=15cm]{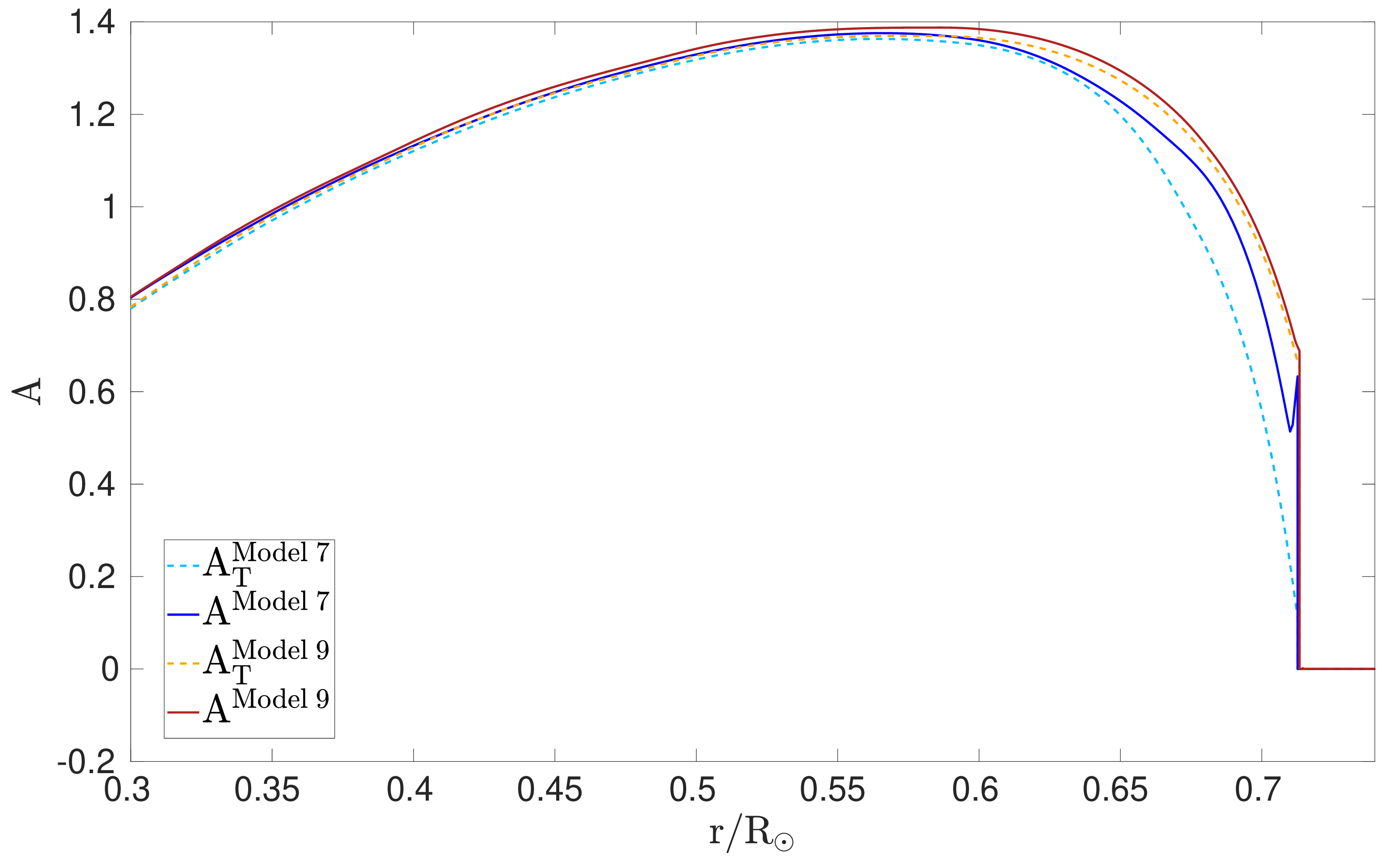}
	\caption{\textbf{Ledoux discriminant profile and the thermal components for two evolutionary solar models.} The dark blue solid line is associated with a standard solar model (Model 7) while the red plain line is for a model including macroscopic transport of chemicals at the BCZ (Model 9). The thermal component of the Ledoux discriminant for both models is plotted as light blue and orange dashed lines for the standard and non-standard evolutionary models respectively. The difference between the dashed and the plain line is, by definition, the contribution of the mean molecular weight term. Source Data provided with this paper.}
		\label{FigADecomp}
\end{figure*} 

\subsection*{Comparisons to theoretical opacity computations}

After reconstruction, we get a ``seismic'' determination of the solar opacity between $0.35\rm{R_{\odot}}$ and the base of the convective zone, providing a data-driven mean Rosseland opacity profile in the solar radiative zone, complementary to experimental values \cite{Bailey,Hoarty2024}. We use it in detailed comparisons with theoretical computations, as shown in Figure 3 for two models of our set, namely Models 1 and 2 (See Table 1), using either the AGSS09 (left panel) or AAG21 (right panel) abundances. We compare our results to the most recent versions of two opacity computation codes, OPAS \cite{Blancard2012} (left panel)  and SCO-RCG (right panel) \cite{Pain2015} (see Methods, subsection Theoretical opacity computations). The ``seismic'' opacity shows for all models a localized increase of $\approx10\%$, at the position of the iron opacity peak at the BCZ. Tests investigating the impact of radiative accelerations show that they are insufficient to explain this increase (see Methods, subsection Parametrization of the solar core). 

Detailed computations have been performed for the exact thermodynamic conditions of both the evolutionary model, before reconstruction, denoted ``Ref'' and the seismic model, denoted ``Seismic''. Results for evolutionary models are shown in brown and in orange for seismic models. The plain lines denote results from the models and the dashed ones those of detailed theoretical computations for the exact same thermodynamic coordinates, namely T, $\rho$ and chemical composition. In the OPAS case (left panel), the dashed and plain brown lines show the consistency between the 2015 and 2021 versions. The SCO-RCG results are compared to the OPAL tables used in the solar model. They are higher than OPAL values by about $9\%$ at $0.35R_{\odot}$ and about $35\%$ at the BCZ for the exact conditions of the model.

\begin{figure*}
	\centering
		\includegraphics[width=15.8cm]{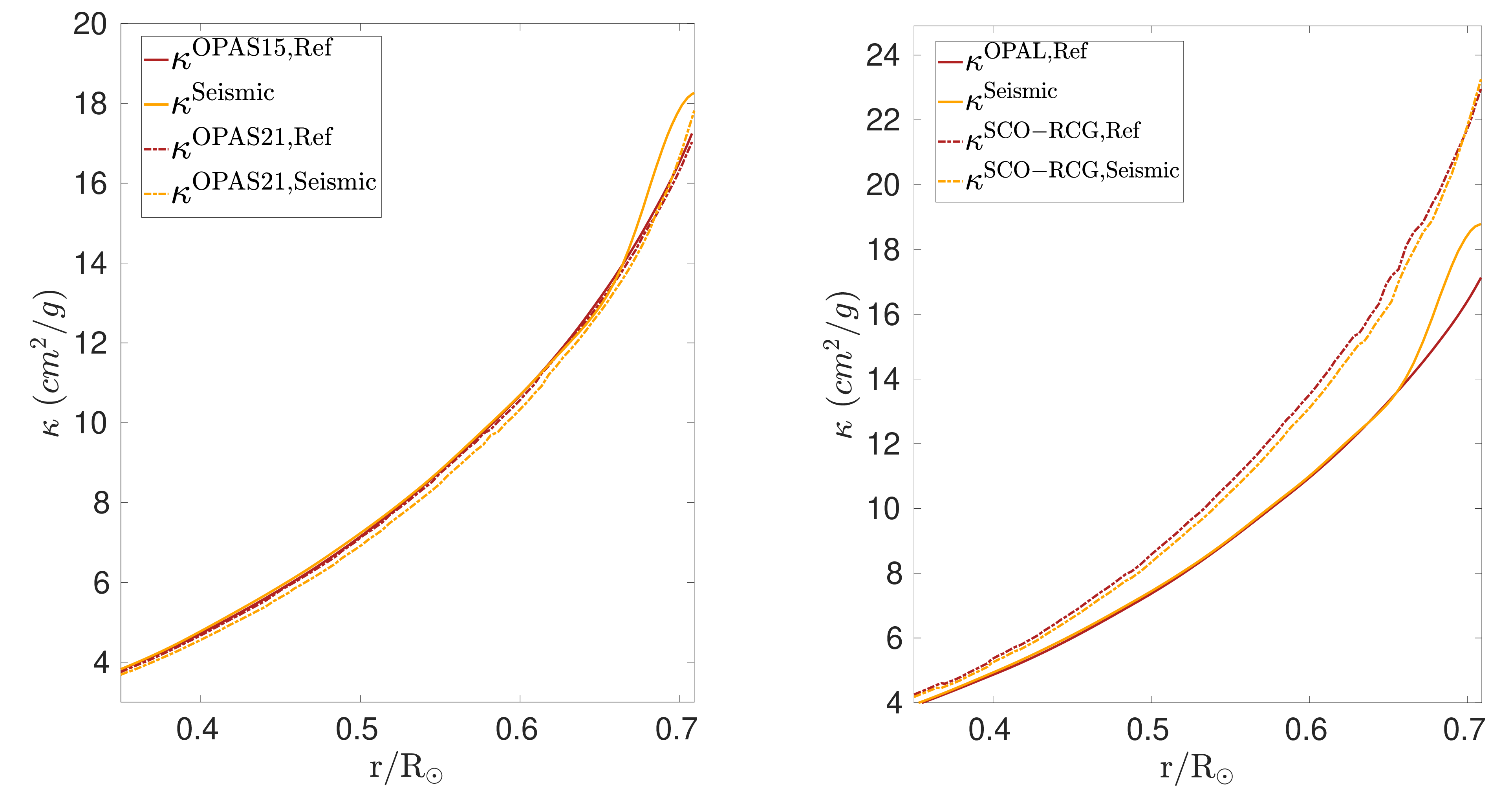}
	\caption{\textbf{Comparison of the mean Rosseland opacity profiles of reference evolutionary models, seismic reconstruction and the updated theoretical computations associated with the thermodynamical paths of both models.} \textit{Left panel:} Opacity profiles from evolutionary computations with the 2015 OPAS tables (brown-plain line), from seismic reconstruction (orange-plain line), and from detailed computations with the OPAS code for both evolutionary and seismic thermodynamical paths (brown and orange dashed lines, respectively), using the AGSS09 abundances. \textit{Right panel:} Opacity profiles for evolutionary computations using OPAL tables (brown-plain line), from seismic reconstruction (orange-plain line) and from detailed SCO-RCG computations for both evolutionary and seismic thermodynamical paths (brown and orange dashed lines, respectively), using the AAG21 abundances. Source Data provided with this paper.}
		\label{FigOpacComp}
\end{figure*}

For both seismic models, the opacity from theoretical computations is reduced by about $2\%$ compared to the evolutionary results (comparing the dashed orange line to the dashed brown one), due to the change in $\rho$ and T from the reconstruction. OPAS computations show a lower opacity than the seismic value by about $10\%$ at the BCZ, while SCO-RCG results show excess of approximately $6\%$ at lower radius (thus higher temperature) and about $24\%$ at the BCZ. 

To investigate the physical origin of the changes, we look at the contributions of the most important elements at the BCZ, namely iron, neon and oxygen for SCO-RCG, OPAS21 (the latest version of the code), OPLIB and OP (version 3.3), the latter being used for the latest standard solar models\cite{Vinyoles2017}. These results are shown in Figure 4 for the exact same thermodynamical conditions of the seismic model as in Figure 3.

\begin{figure*}
	\centering
		\includegraphics[width=15.5cm]{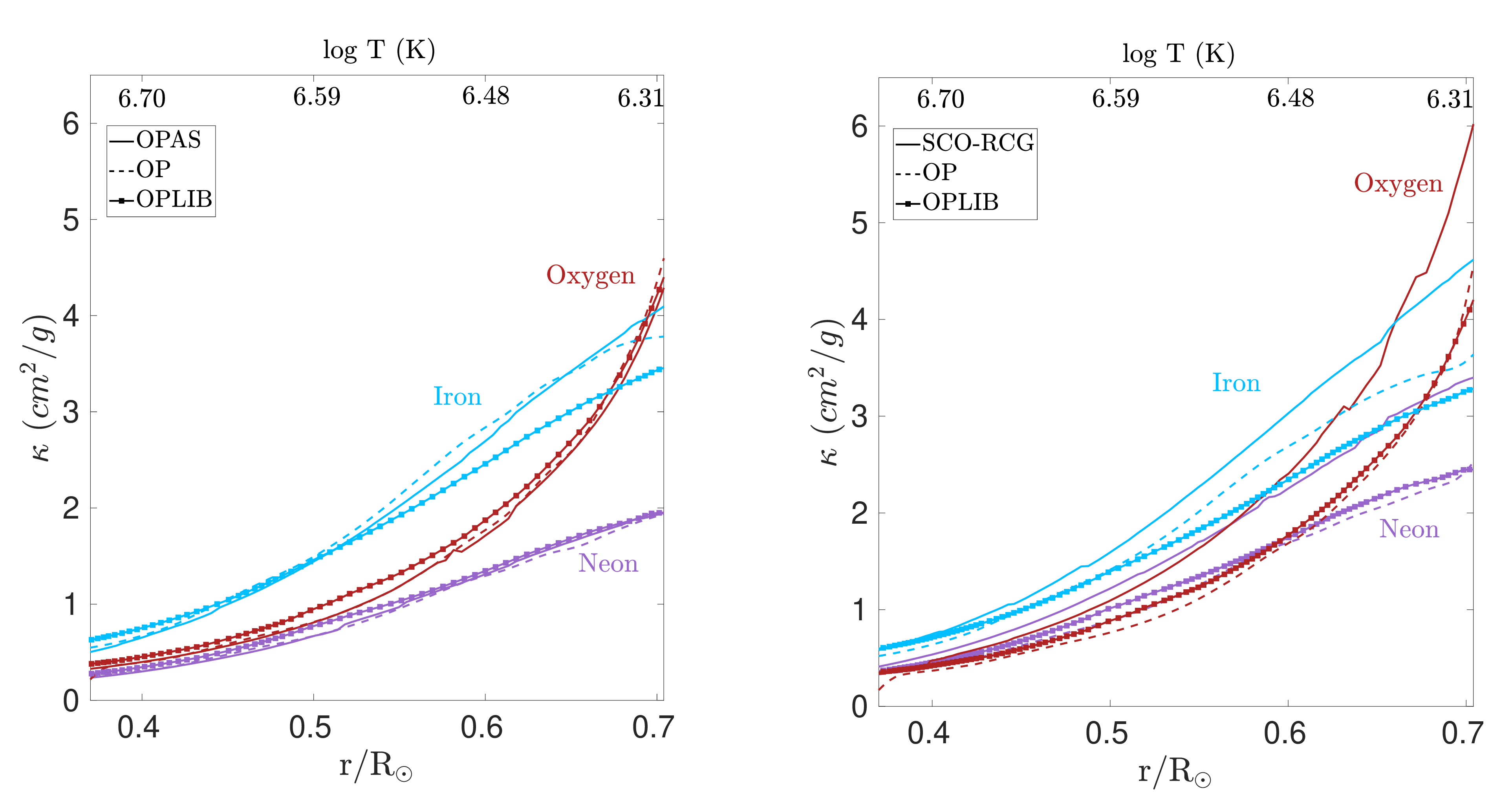}
	\caption{\textbf{Opacity profiles in the bulk of the radiative zone for the three main contributors to opacity at the BCZ, namely oxygen, neon and iron for the seismic thermodynamical path.}  \textit{Right panel:} SCO-RCG, OPLIB and OP values using AAG21 abundances. \textit{Left panel:} \textbf{OPAS21}, OPLIB and OP values using AGSS09 abundances. Source Data provided with this paper.}
		\label{FigOpacContribMain}
\end{figure*} 

Comparing the left and right panels in Figure 4 shows the impact of the neon abundance revision, noted previously \cite{Buldgen2019, Zhang2019}. For other elements, we compare in the right panel OPLIB, OP and SCO-RCG, that used the exact same thermodynamical conditions. We find that SCO-RCG always shows significantly higher opacities than both OP and OPLIB, neon and oxygen showing large increases even at high temperatures. Thus, a model using SCO-RCG opacities might show a high helium abundance and central temperature, altering the predicted neutrino fluxes. Similar changes are seen for other contributors such as silicon, magnesium, carbon and nitrogen (See Methods, subsection Theoretical opacity computations). The origin of these variations is found in the treatment of the Stark effect for the lighter elements, whereas additional plasma effects, the number of transitions and energy levels as well as the Partially Resolved Transition Arrays used in the computation impacts the results for heavier elements such as iron (see Methods, subsection Theoretical opacity computations for details of the opacity codes). Comparisons between OPAS, OPLIB and OP opacities show much smaller differences, mainly for iron and oxygen, but remain comparable throughout the model except for the BCZ where larger deviations appear. The case of iron in OPLIB is remarkable as the deviations start at relatively high temperature. This trend follows the one seen for heavier elements in the OPLIB computations. As the oxygen, neon and iron revision by Magg et al. (2022)\cite{Magg2022} is at the origin of the improved sound speed agreement of their standard solar models using OP opacities, our results show that this agreement is unlikely to hold if other opacity tables were used in their calibration of the solar models. This statement is confirmed by the seismic opacity determined for Model 10, discussed in the Methods section. Hence, the differences between OP, OPLIB, OPAS and SCO-RCG in the same thermodynamical conditions call for further investigations before the solar modelling problem can be considered solved.
\section*{Discussion}
Opacities have long been seen as an important source of disagreements between theoretical solar models and helioseismic data. Following the recent experimental results \cite{Bailey, Nagayama2019,Hoarty2024} and the debate related to solar surface abundances, we provide a ``seismic opacity'', i.e. stringent constraints on solar opacity by combining seismic reconstruction techniques based on the solar models of Eggenberger et al. (2022)\cite{Eggenberger2022} reproducing lithium, helium and the internal solar rotation. Our inference results highlight an opacity shortage of about $10 \pm 2\%$ at the BCZ in solar models with OPAL, OPLIB, OPAS and OP tables, in line with the experimental results \cite{Bailey,Hoarty2024}. Comparing our values to detailed opacity computations in the exact same solar conditions, we find differences of up to $40\%$ between the SCO-RCG code and the OPAS and OP codes, as well as disagreements with helioseismic values. We demonstrate the power of helioseismology in using the Sun as a fundamental physics experiment, as was done for the equation of state in the past \cite{JCD88,Elliott1998}, tightly constraining here the behaviour of radiative opacities, a key ingredient of stellar models and a major source of uncertainty. The differences between the various theoretical opacities we observe somewhat recall the issues found for Cepheids in 1980s and we conclude with a plea similar to that of Simon (1982) and Andreasen $\&$ Petersen (1988)\cite{Simon1982,Andreasen1988} and of recent asteroseismic studies of massive stars \cite{Moravveji2016, DD2017}. The need for new computations of detailed opacity tables and analyses of the treatment of physical processes and numerical techniques demonstrate the key role of helio- and asteroseismology to guide these works. We show that opacities constitute a major contributor to the solar and stellar modelling problem. Full evolutionary computations with SCO-RCG and OPAS tables will impact the way we see the Sun and stars in the Universe as well as stellar populations accross cosmic times. 

\newpage

\renewcommand{\figurename}{Supplementary Figure}

\setcounter{figure}{0}

\section*{Methods}\label{Sec:Methods}

\subsection*{Solar evolutionary models and macroscopic transport of chemical elements}

The starting point of the reconstruction procedure is provided by calibrated solar evolutionary models computed with the Li\`ege stellar evolution code (CLES) \cite{ScuflaireCles}. The calibration procedure uses four free parameters, namely the initial hydrogen mass fraction, $X_{0}$, the initial metallicity, $Z_{0}$, the mixing-length parameter of convection, $\alpha_{\rm{MLT}}$, and an envelope overshooting parameter, $\alpha_{\rm{Ov}}$, optimized to reproduce the present-day solar luminosity, $L_{\odot}$, the solar radius, $\rm{R_{\odot}}$, the surface heavy element abundance $\left(Z/X\right)_{\rm{S}}$ as well as the position of the base of the solar convective zone, $r_{CZ}$. The properties of the set of calibrated evolutionary models are listed in Table 1. The illustrations (Supplementary Figures) of all the tests carried out in the method section are presented in the Supplementary Information. 

Our method was applied to an extensive set of reference models, varying ingredients at microscopic and macroscopic scales. Such an approach is required to analyze the effect of the solar calibration procedure, and of other factors such as abundances, transport of chemicals, etc., to determine whether the trends we have seen for Models 1 and 2 will not be erased by a specific combination of physical ingredients. We examined 10 solar seismic models from 10 individual calibrations. Regarding microscopic physics, we relied on using the AGSS09, AAG21, MB22 and the GN93 abundances. Other variations included the equation of state (FreeEOS \cite{Irwin}, SAHA-S \cite{Gryaznov04,Baturin2013}), radiative opacity tables (OP \cite{Badnell2005}, OPLIB \cite{Colgan}, OPAS \cite{Mondet}, OPAL \cite{OPAL}), and nuclear reaction rates \cite{Adelberger,NacreI}.  Microscopic diffusion is taken into account in all calibrations following Thoul et al. (1994)\cite{Thoul}, including the screening coefficients of Paquette et al. (1986)\cite{Paquette} and the effects of partial ionization. Regarding macrophysics, we tested various formulations of the empirical coefficients of Proffitt $\&$ Michaud (1991) \cite{Proffitt1991}, including the recent recalibrations presented in Eggenberger et al. (2022)\cite{Eggenberger2022} to reproduce the combined effect of hydrodynamic and magneto-hydrodynamic instablities \cite{Eggenberger2005,Eggenberger2019} in the non-rotating CLES models and a standard model (for the GN93 abundances), not including any of such effects. 

The formalism for macroscopic transport in Eggenberger et al. (2022)\cite{Eggenberger2022} is based on the transport of angular momentum in stellar radiative zones under the hypothesis of shellular rotation, combining the shear instability, the meridional circulation, and the Tayler-Spruit dynamo \cite{Spruit2002}. In such conditions, the vertical transport of chemical elements follows a diffusion equation 
\begin{align}
\frac{\partial X_{i}}{\partial t}=\frac{1}{\rho r^{2}}\frac{\partial}{\partial r}\left[\rho r^{2} D_{X} \frac{\partial X_{i}}{\partial r}\right],
\end{align}
with $\rm{X}_{i}$ a given chemical species, $\rho$ the local density, $\rm{r}$ the radial position of a given isobar and $\rm{D_{X}}$ a diffusion coefficient that accounts for the impact of the additional physical processes. The shear diffusion coefficient we consider is from Talon $\&$ Zahn (1997)\cite{Talon1997} and includes the effects of the stabilizing mean molecular weight gradient in the analytical expression. This coefficient is written
\begin{align}
\rm{D_{X}}\approx \frac{2 Ri_{c}(dU/dz)^{2}}{N^{2}_{T}/(K+D_{h})+N^{2}_{\mu}/D_{h}}, \label{eq:ShearTalon}
\end{align}

with $\rm{D_{h}}$ the horizontal turbulence coefficient, $Ri_{c}$ the critical Richardson number, $dU/dz=r \sin\theta (\rm{d}\Omega/\rm{d}r)$, the shear rate, K the thermal diffusivity and $N_{\mu}=\frac{g}{r}A_{\mu}$ and $N_{T}=\frac{g}{r}A_{T}$ the chemical and thermal contribution to the Brunt-V\"ais\"al\"a frequency. 

The asymptotic formulation for the diffusion coefficient is derived by assuming that the denominator can be simplified to the sole contribution of the mean molecular weight gradient. Namely that $\rm{N^{2}_{T}/(K+D_{h})}\ll \rm{N}^{2}_{\mu}/\rm{D_{h}}$ and that the shear rate will be determined by the condition for the Tayler-Spruit instability to set in, in regions where mean molecular weight gradients dominate. This condition states that
\begin{align}
\bigg| \frac{\rm{d} \ln \Omega}{\rm{d} \ln r}\bigg| \geqslant \left(\frac{N_{\mu}}{\Omega}\right)^{7/4}\left( \frac{\eta}{r^{2}N_{\mu}}\right)^{1/4}.\label{eq:crit}
\end{align}
$\Omega$ is assumed here as a typical angular velocity in the radiative zone (fixed to the helioseismic value), $\eta$ is the magnetic diffusivity. Introducing this criterion in Eq. \ref{eq:ShearTalon}, assuming equality in Eq. \ref{eq:crit} (i.e. that the value of the shear is kept just at the critical level for the instability to operate), and averaging over latitude, one can derive the following expression

\begin{align}
    D_{X,1}=D_{h}f(r)\Omega^{-3/2} \left(\frac{\eta \vert N^{2}_{\mu} \vert}{r^{2}}\right)^{1/2}, \label{Eq:DiffCoeff}
\end{align}

where  $\rm{f(r)}$ is an appropriate weight function used to reproduce the more complex behaviour of the mixing coefficient. The magnetic diffusivity is approximated considering solar material as a hydrogen plasma

\begin{align}
\eta \approx 5.2 \times 10^{11} \left(\frac{\ln \Lambda}{T^{3/2}}\right),
\end{align}
with $\Lambda =-12.7+ \log \rm{T} -0.5 \log \rho$, the ratio of the Debye length to the impact parameter in the plasma of electron concentration, assuming cgs units.

The coefficient represents the interaction of meridional circulation, shear-induced turbulence, and the Tayler-Spruit instability. Both $f(r)$ and $\Omega$ are adapted for the solar case to replicate results from models with full magneto-hydrodynamical treatment of rotation in the Geneva stellar evolution code \cite{Eggenberger2008}. $D_{h}$ serves as a free parameter for calibrating transport efficiency, representing the value attributed to horizontal shear-induced turbulence.

Another common approach is to implement a simple coefficient depending only on the local density, following \cite{Proffitt1991}

\begin{align}
D_{X,2}=D_{T}\left(\frac{\rho}{\rho_{\rm{BCZ}}}\right)^{-n}, \label{Eq:DiffProff}
\end{align}

with $\rho_{\rm{BCZ}}$ the density at the lower border of the convective envelope and $\rm{D_{T}}$ and $n$ are free parameters. 

Both parametrizations reproduce solar-age lithium depletion and maintain a helium value in the convective envelope consistent with helioseismic determinations \cite{Eggenberger2022}. As discussed in Eggenberger et al. \cite{Eggenberger2022}, the overall trend is independent of the coefficient's form. Internal gravity waves \cite{Charbonnel2005,Rogers2005} have also been proposed for solar rotation profile flattening and lithium depletion. In this scenario, lithium depletion occurs through shear layer oscillation, modeled as a diffusive coefficient in stellar evolutionary models \cite{Charbonnel2005}. The wide range of coefficients studied in \cite{Eggenberger2022} and \cite{Buldgen2023} likely captures this effect. However, models with effects of internal gravity waves, as per \cite{Charbonnel2005} and \cite{Pincon2016}, would be interesting to investigate. In this study, Eq. \ref{Eq:DiffCoeff} is used for Models 1 to 6, and Eq. \ref{Eq:DiffProff} for Model 8. Model 7 is a standard solar model without macroscopic chemical element transport, and Model 9 has lower macroscopic mixing efficiency by about 30$\%$, influencing the light element depletion defined by Eq. \ref{Eq:DiffCoeff}, reproducing solar lithium depletion within one sigma using a lower $D_{h}$ value.

The mixing of chemical elements in the overshooting region is assumed instantaneous and the temperature gradient to be the adiabatic gradient. Thus $\alpha_{\rm{Ov}}$ simply extends deep enough the convective envelope so that the temperature gradient transition is placed at the position inferred from helioseismology. This is achieved with an extended calibration procedure using $4$ parameters instead of the usual $3$ used for standard solar models. These use two parameters describing the initial chemical composition and one describing the efficiency of convection for standard calibrations, aiming at reproducing the solar radius, luminosity and surface metallicity at the current solar age. Here, we add an additional parameter for the efficiency of convection to recover the base of the convective zone.

The reconstruction method uses an iterative correction of the Ledoux discriminant profile defined in Eq. 1. Since $A^{\mu}\ll A^{T}$ when macroscopic mixing occurs at the BCZ, we essentially have a direct measure of the temperature gradients from the inversion of the Ledoux discriminant. Essentially, efficient mixing of the chemical elements renders the Ledoux discriminant to the Schwarzschild discriminant. The method is similar to that of Baturin et al. (2015)\cite{Baturin2015} to study the composition profile at the BCZ, or that of Gough (2004)\cite{Gough2004} to determine the opacity of a seismic model. 

The reconstruction procedure yields density and pressure profiles consistent with helioseismic data. As shown in Buldgen et al. (2020)\cite{Buldgen2020}, an agreement of around $0.1\%$ is achieved for all structural inversions after the reconstruction. Using the obtained pressure ($P$) and density ($\rho$) along with the chemical composition profile of the non-standard evolutionary models, defined by the hydrogen (X) and heavy elements (Z) abundance, we infer the temperature $T=T(\rho, P, X, Z)$ at each point of the radiative zone using any equation of state available for classical stellar evolutionary computations. We have assumed that the transport of chemical elements has been properly taken into account during the solar history by the non-standard models fitting the surface metallicity, the lithium and helium abundances at the age of the Sun.

\subsection*{Parametrization of the solar core}

Buldgen et al. (2020)\cite{Buldgen2020} do not consider thermal equilibrium in solar models. As they rely on adiabatic oscillation equations, direct constraints on temperature and chemical composition are not provided. For more information and additional references on seismic solar models, we refer to \cite{Shibahashi1995,Basu1996,Turck2004}. Assuming a specific equation of state enables the inference of ``secondary'' thermodynamic variables like temperature or chemical composition (See e.g., \cite{Shibahashi1993,Antia1995,Roxburgh1996,Kosovichev1996,Antia1997}). This study aims to reconstruct a full opacity profile from seismic inversions using the radiative transfer equation. This is meaningful only if the energy production equation is satisfied—meaning the right amount of energy is produced in the solar core by nuclear reactions—and if thermal equilibrium is assumed for seismic solar models. To achieve this, we ensure that the luminosity at $0.3\rm{R_{\odot}}$ matches the solar luminosity. Inspections of solar models show a plateau in the luminosity profile at $0.3\rm{R_{\odot}}$ that corresponds to the surface value within $0.1\%$. Physically, this implies that above this limit, energy generated in the core by nuclear reactions is simply transported outward.

Achieving this involves parametrizing the chemical composition profile below $0.3\rm{R_{\odot}}$ to find a solution matching the energy generation of the evolutionary model. The chemical profile's parametrization is discussed below, and the minimization procedure is rigorously constrained to avoid solutions with unphysical chemical composition gradients and significant deviations from the evolutionary model. At $0.3\rm{R_{\odot}}$, the chemical composition profile precisely aligns with that of the evolutionary model. Consequently, above this limit, it remains unaffected by core region parametrization and is solely determined by evolutionary computations.

This second reconstruction step is based on equation
\begin{equation}
\frac{dL}{dr}=4 \pi r^{2}\rho \epsilon(\rho,T,X,Z), \label{Eq:NuclearGen}
\end{equation}
with L the luminosity, $\rho$ the density of the seismic model, X and Z coming from the parametrization, T the temperature profile given by the equation of state and $\epsilon$ is the energy generation rate of nuclear reactions, which is computed from the nuclear energy generation routines of the Li\`ege stellar evolution code. The contribution from gravity, $\epsilon_{\rm{g}}$ is less than $0.1\%$ and thus negligible.

As noted by Gough (2004)\cite{Gough2004}, the reproduction of the solar luminosity by the seismic model depends on the amount of helium in the core. Thus, the chemical composition of the core is iterated so that the model reproduces the luminosity plateau observed around $0.3\rm{R}_{\odot}$. Previous approaches in Gough $\&$ Scherrer (2001)\cite{Gough2001}, Gough (2004)\cite{Gough2004} changed the global core helium content, or assumed a constant metallicity in the solar interior\cite{Shibahashi1996,Shibahashi2006} to provide a unique solution to the reconstruction of the solar thermal structure. Here, we use the following parametrization of the chemical profiles to that end
\begin{eqnarray}
\rm{X(\rm{r})}&=&\rm{X_{0}}(\rm{r_{Sup}}) +\left(\alpha_{3}X_{0}(0)-X_{0}(\rm{r_{Sup}})\right)\left(\frac{(X_{0}(\rm{r})-X_{0}(\rm{r_{Sup}}))}{(\alpha_{3}X_{0}(0)-X_{0}(\rm{r_{Sup}}))}\right)^{\alpha_{1}}\exp{(-\alpha_{2}(r/R_{\odot})^{2})},  \\
\rm{Y(\rm{r})}&=&\rm{Y_{0}}(\rm{r_{Sup}}) +\left(\alpha_{4}Y_{0}(0)-Y_{0}(\rm{r_{Sup}})\right)\left(\frac{(Y_{0}(\rm{r})-Y_{0}(\rm{r_{Sup}}))}{(\alpha_{4}Y_{0}(0)-Y_{0}(\rm{r_{Sup}}))}\right)^{\alpha_{1}}\exp{(-\alpha_{2}(r/R_{\odot})^{2})} ,\\
\rm{Z(\rm{r})}&=&1.0-\rm{X(\rm{r})}-\rm{Y(\rm{r})},
\end{eqnarray}
with the free parameters $\alpha_{i}$. For sufficiently small variations of the free parameters, this approach preserves the properties of chemical composition gradients and ensures that $X+Y+Z=1$ as well as a continuous connection at $r_{Sup}$ (here $0.3R_{\odot}$). $X_{0}(r)$ and $Y_{0}(r)$ denote the hydrogen and helium mass fraction profiles of the evolutionary model. The parameters will take values close to $1$, with multiple solutions being possible. Usually, $\alpha_{1}$ will be about $0.98$, $\alpha_{2}$ about $1$, and $\alpha_{3}$ and $\alpha_{4}$ only vary by about $0.006$.

The parametrization proposed in equations 9-11 is not unique and similar reconstruction techniques applied to other evolutionary stages or masses would require modifications. Such methods could in principle be applied to other types of pulsators, provided that the inversion technique is adapted to handle the lower number of oscillation modes. The parametrizations of regions where the mean molecular weight gradient dominates the Ledoux discriminant profile would then be highly informative on the transport of chemicals.

Tests on solar model profiles have demonstrated its effectiveness in replicating trends observed in evolutionary computations, as depicted in Supplementary Fig. 1. Notably, recent work by Kunitomo $\&$ Guillot \cite{Kunitomo2021} has suggested that considering the solar system's formation could lead to a local increase in the solar core's metallicity. Such considerations should be integrated into upcoming seismic solar models. Kunitomo $\&$ Guillot \cite{Kunitomo2021} also confirmed the necessity for increased opacity to enhance agreement between solar models and helioseismic results, even when accounting for planetary formation, accretion, and mass loss. Attempting to replicate solar neutrino flux observations may prove challenging, given the potential impact of evolutionary history on model predictions\cite{Kunitomo2022}. As demonstrated below, our results remain insensitive to events within the solar core.

The optimization is carried out with a Levenberg-Marquardt algorithm and converges after about 5 iterations on a parametric profile reproducing the solar luminosity within $0.1\%$. The Levenberg-Marquardt method uses as constraints the luminosity profile of the reference model between $0.1$ and $0.3\rm{R}_{\odot}$ on about 30 points. Fitting directly the nuclear energy generation rate, $\epsilon(\rho, T, X, Z)$, also allows to recover accurately the opacity profile.  

\begin{figure*}
	\centering
		\includegraphics[width=15.5cm]{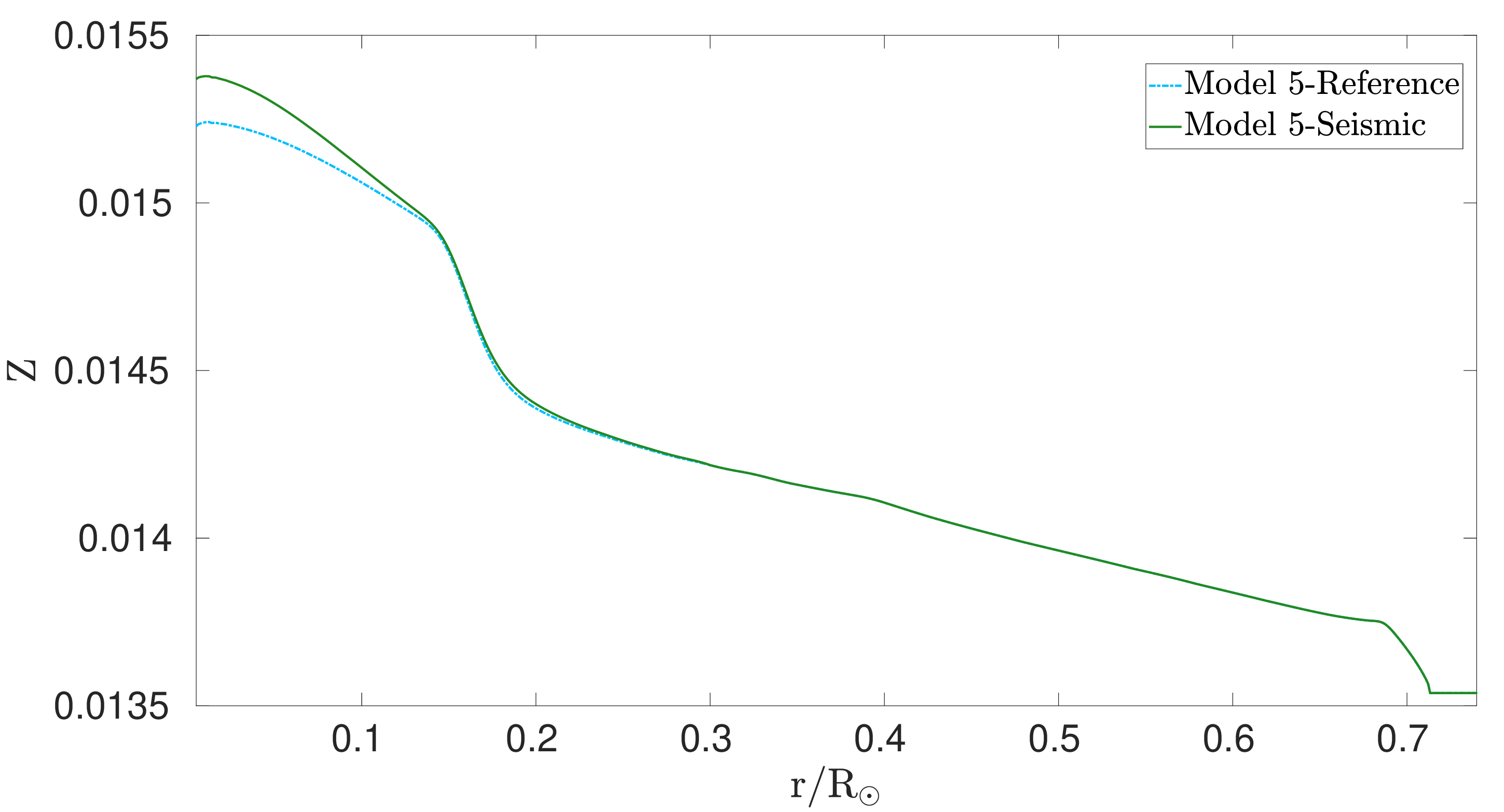}
	\caption{Heavy element mass fraction profile (Z) as a function of normalized radius $\rm{r/R_{\odot}}$ in the radiative zone of a reference model and its seismic counterpart (here taken for Model 5 that includes macroscopic transport at the BCZ and the OPLIB opacities as reference radiative opacities in the calibration).}
		\label{FigZParam}
\end{figure*} 


Once the luminosity profile is recovered, we compute the mean Rosseland opacity, $\kappa$, from the radiative transfer equation

\begin{equation}
\kappa = -\frac{16\pi r^{2} a_{\rm{rad}}cT^{3}}{3 \rho L}\frac{dT}{dr}, \label{Eq:radTransfer}
\end{equation}
with c the speed of light in vacuum and a$_{\rm{rad}}$ the radiation density constant.

The stability of the procedure is tested by recovering the opacity profile from an evolutionary model using the parametrized core profile. We found that if the luminosity is well reproduced, the opacity is recovered within less than $0.1\%$, which is suitable for comparisons with theoretical computations.

Attributing all corrections in temperature gradient at the BCZ to opacity modifications implies that this region is purely radiative. This might not be entirely the case and changes could be attributed to the opacity while in reality stemming from the thermalization of convective elements. Monteiro et al. (1994)\cite{Monteiro94} showed that some additional penetration at the BCZ could improve the agreement with helioseismic data.

This introduces uncertainties in our inferred opacity values. However, based on recent results\cite{JCD2011,Zhang2019}, this transition should not extend below $0.68 \rm{R_{\odot}}$. Besides the temperature gradient transition, it is central that mixing reproduces the current solar lithium and beryllium abundances, along with lithium abundances observed in young solar twins in open clusters. In Zhang et al. (2019)\cite{Zhang2019}, the chemical mixing reaches nearly $0.6\rm{R}{\odot}$ for a temperature gradient transition at $0.68 \rm{R{\odot}}$. Therefore, if the observed temperature gradient transition in our inversions was only due to overshooting, it would lead to much deeper mixing and excessive beryllium and lithium depletion, especially during the pre-main sequence. While this does not entirely rule out overshooting, it suggests it should not be the sole explanation, leaving opacity as the only other candidate.

Another point is the influence of radiative accelerations, extensively studied in solar models \cite{Turcotte1998, Schlattl2002, Gorshkov2008}. Their impact on solar structure is minimal, primarily near the BCZ. However, a significant increase in iron opacity would amplify the effect of radiative accelerations for this element. This, in turn, would strengthen iron's contribution at the BCZ, but this impact would be counteracted by the efficient macroscopic mixing needed to reproduce the lithium depletion. Therefore, it is unlikely that radiative accelerations play a major role in recovering the missing opacity at the BCZ. To confirm this, we tested their impact under conditions maximizing their effect—calibrated standard solar models without macroscopic mixing using AAG21 abundances. We found the impact to be around $0.5\%$, insufficient to explain the observed modifications. This conclusion, however, applies specifically to the solar case, as demonstrated in Deal et al. (2018)\cite{Deal2018}.

All evolutionary models undergo the seismic reconstruction phase individually, following a similar procedure to Buldgen et al. (2020)\cite{Buldgen2020}. This ensures complete consistency and provides a better understanding of the result dispersion after seismic reconstruction. For completeness, tests using various helioseismic datasets \cite{Larson2015} have been conducted, to quantify the uncertainties in the final opacity profiles stemming from helioseismic data. In Supplementary Fig. 2, we present final reconstructed Ledoux discriminant profiles for various test cases with different physics. The profile is uniquely determined from the inversion, except for the narrow region between $0.7$ and $0.713$ solar radii and the reconnection point around $0.1$ solar radii, where some variations are observed. Supplementary Fig. 3 illustrates the chemical composition profiles at the BCZ for some models entering the opacity reconstruction procedure. Models in Supplementary Fig. 3 were chosen to illustrate the largest discrepancies in metallicity among the set.

\begin{figure*}
	\centering
		\includegraphics[width=15.5cm]{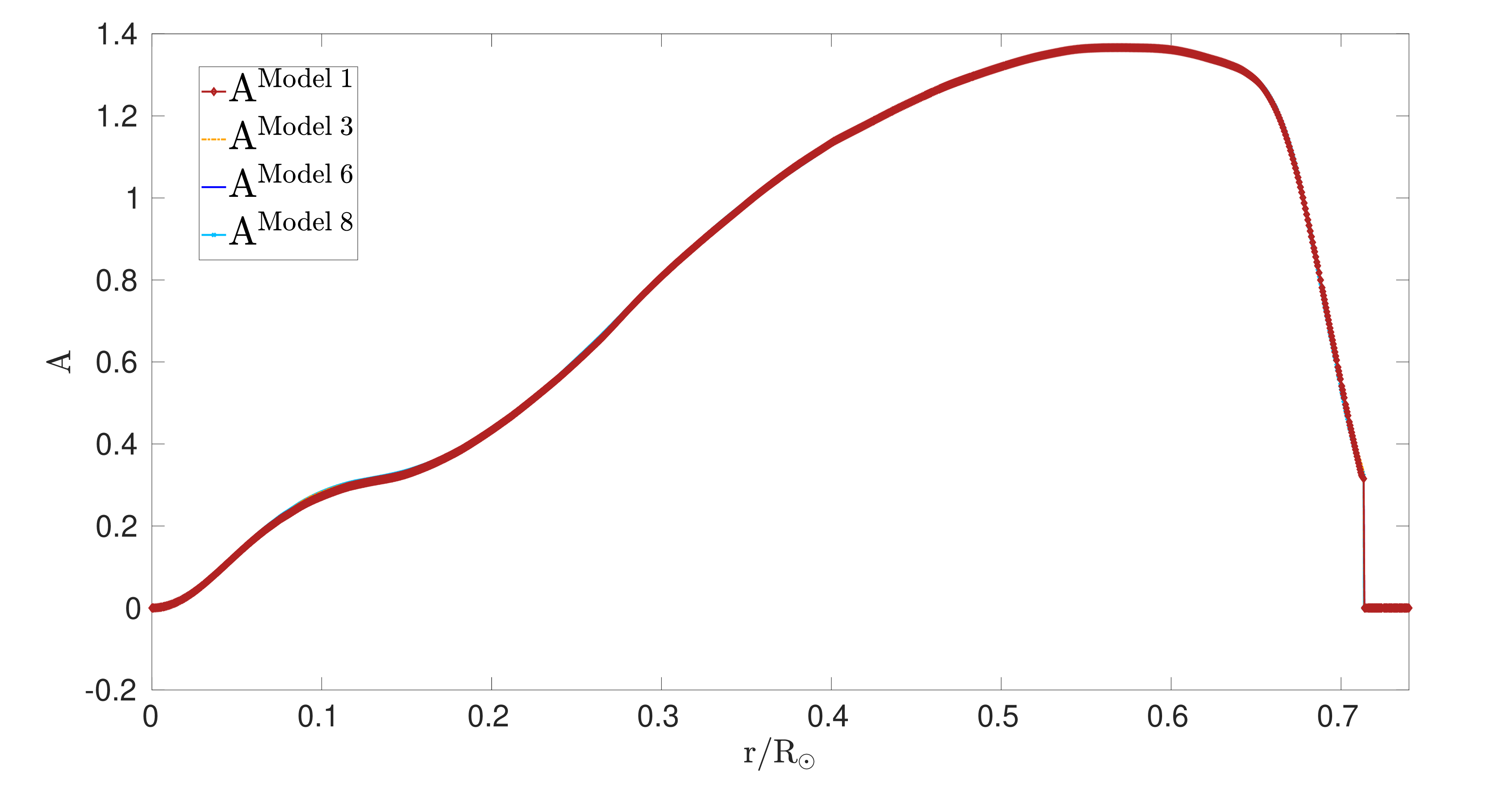}
	\caption{Reconstructed Ledoux discriminant profiles from various initial reference models, varying the reference opacities, chemical abundances and efficiency of macroscopic mixing (See Table \ref{tabAddRefModels} for details of the included physics). The curves are hardly distinguishable, showing the model-independency of the final reconstructed profile.}
		\label{FigAall}
\end{figure*} 

\begin{figure*}
	\centering
		\includegraphics[width=12cm]{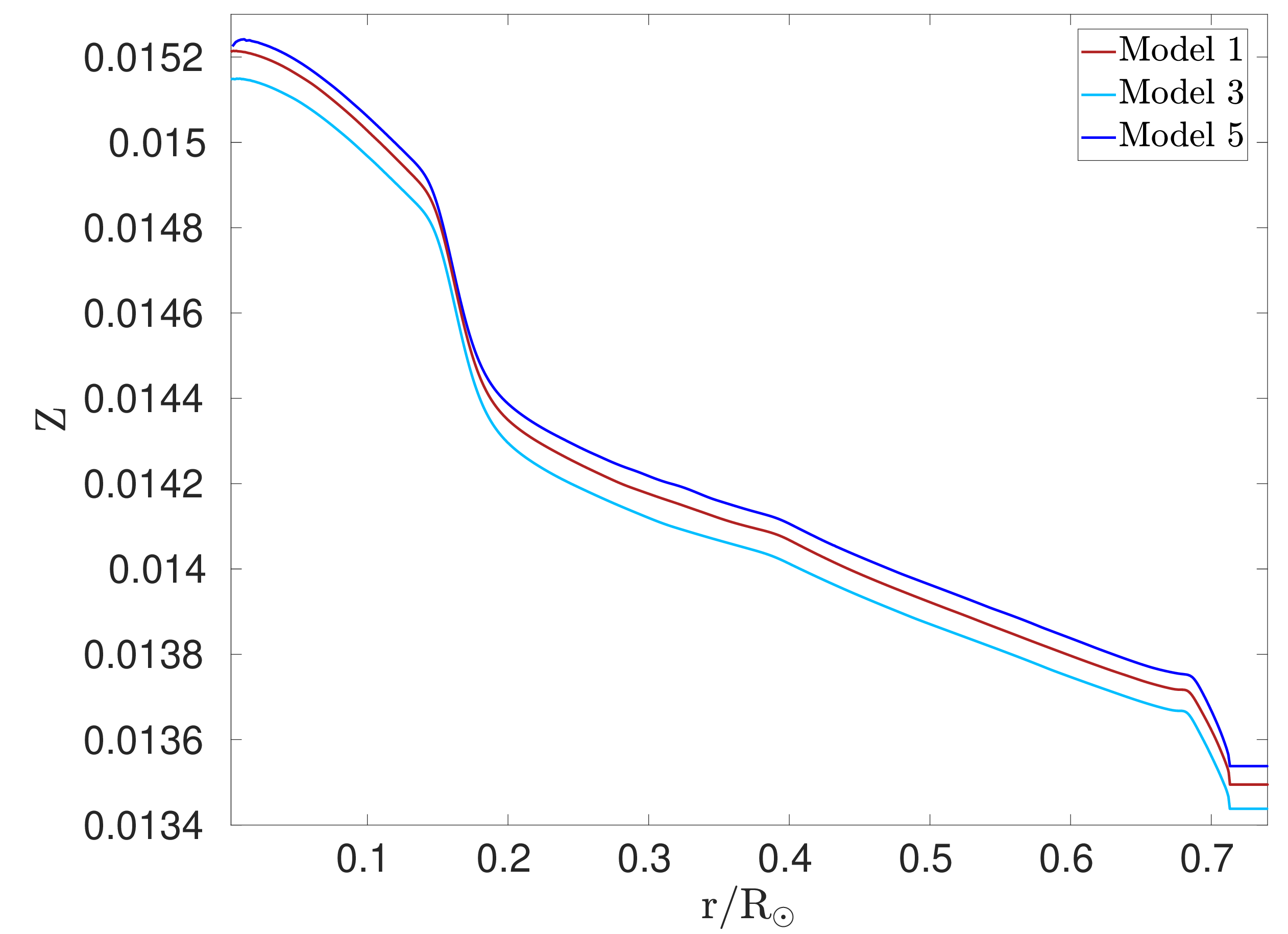}
	\caption{Z profiles for reference Models 1, 3 and 5, illustrating the spread in metallicity values obtained from arious calibrations using different reference opacities, namely OPAS, OP and OPLIB. See Table \ref{tabAddRefModels} for details on the properties.}
		\label{FigZDiff}
\end{figure*} 

This model set is used to test the model dependencies of our opacity reconstruction, while also giving us a larger range of reconstructed profiles to study. These effects are discussed below when illustrating the impact of model uncertainties on the reconstructed opacity profiles. 

\subsection*{Impact of model and dataset uncertainties on the opacity inversion}\label{Sec:ModelUncertainties}

The reconstruction procedure is impacted by a few hypotheses. Some are significant, others do not impact the final result. The first and most important assumption is that the chemical composition profile in the upper radiative layers of the Sun is accurately reproduced by the evolutionary models including macroscopic transport. Changing slightly the transport properties can impact the final result, but does not erase the need for a significant opacity correction at the BCZ. 

Additionally, the necessity for low-metallicity models to replicate the high helium value determined from helioseismology, along with lithium depletion, calls for some form of macroscopic mixing. Various tests with different coefficients are depicted in Supplementary Fig. 4. Two coefficients for macroscopic mixing, $D_{X,1}$ defined in Eq. 5, at two different values of $D_{h}$ varying by about $30\%$ and $D_{X,2}$ defined in Eq. 6, are examined. Additional tests altering the reference opacity tables and equation of state are presented in Supplementary Fig. 5. In every case, these results indicate the need for an opacity correction at the base of the convective envelope, directly resulting from the Ledoux discriminant inversion.

\begin{figure*}
	\centering
		\includegraphics[width=12cm]{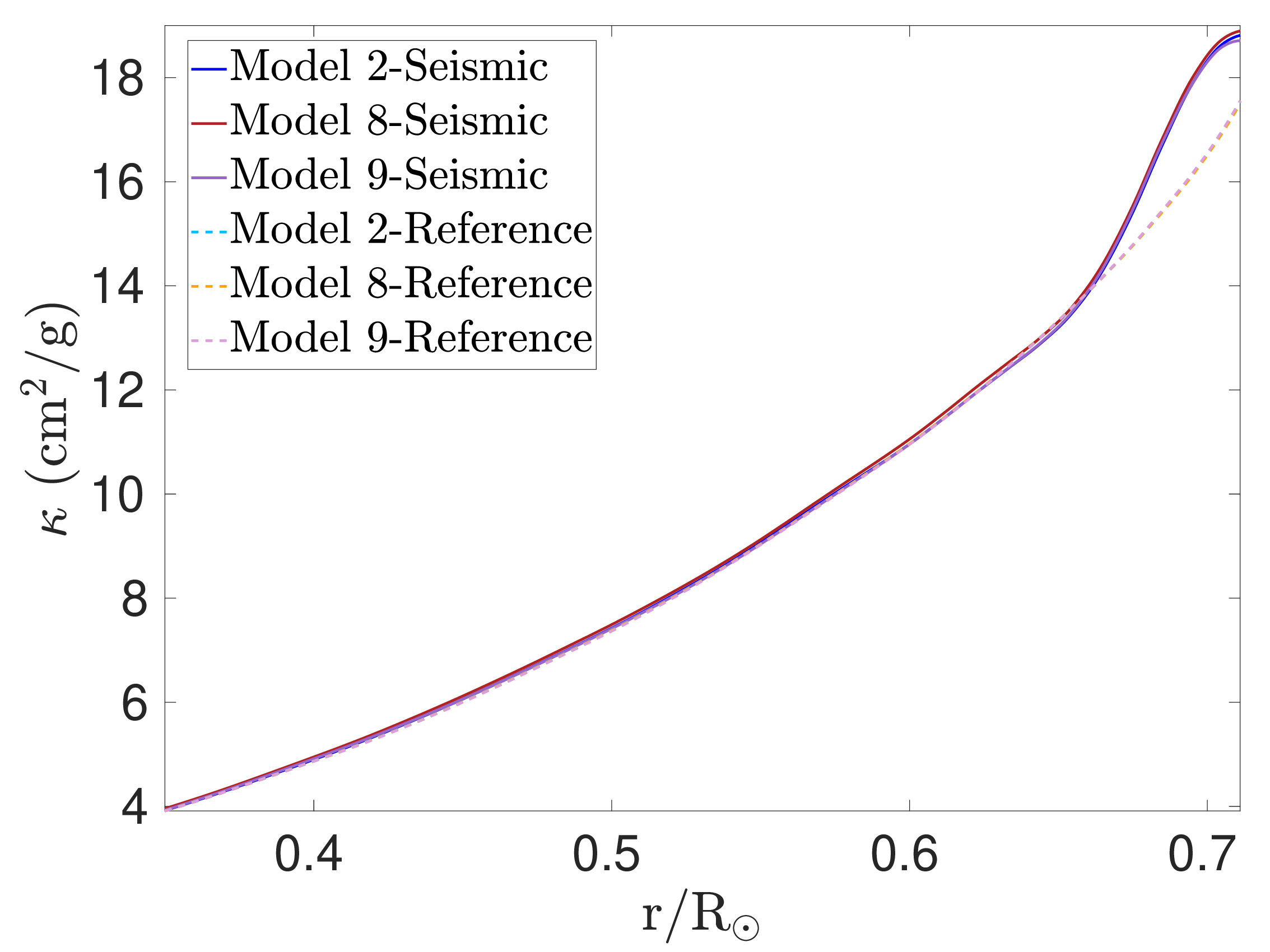}
	\caption{Effects of the mixing parametrization and intensity on the seismic opacity profiles (using Models 2, 8, 9). The reference opacity profiles are shown by the dashed line, while plain lines indicate the corresponding seismic opacities. The dashed curves are almost indistinguishable from each other.}
		\label{FigMixing}
\end{figure*} 

Given their link with the solar problem, we present in Supplementary Fig. 6 a high metallicity standard model, model 7, built using the GN93 abundances and a model built with the recent MB22 abundances. Interestingly, both cases find a significant reduction of opacity around $0.64\rm{R}_{\odot}$, and a small increase closer to the BCZ. This increase is much lower than what is found in AAG21 models, especially in the MB22 model. This indicates that the abundance revision of MB22 leads to a slightly too high opacity around $0.64\rm{R}_{\odot}$ as a result of the higher abundance of metals. This result is quite significant considering that this model includes macroscopic transport reproducing the lithium depletion, which leads to a less steep temperature gradient close to the BCZ than in a standard solar model\cite{Eggenberger2022}. Due to the higher abundance of metals in these models, the overall opacity is higher, particularly at the BCZ. This highlights the fact that the opacity reconstruction is directly linked to the chemical composition of the model and that each unique recalibration may change a bit the scale but not eliminate the fact that a correction is required. This is in line with previous works\cite{JCD2009} that used the opacity of a GN93 model to ``correct'' an AGSS09 model in terms of relative sound speed differences. In that sense, further revision of the opacities will play a key role regarding the solar abundance scale.

\begin{figure*}
	\centering
		\includegraphics[width=15cm]{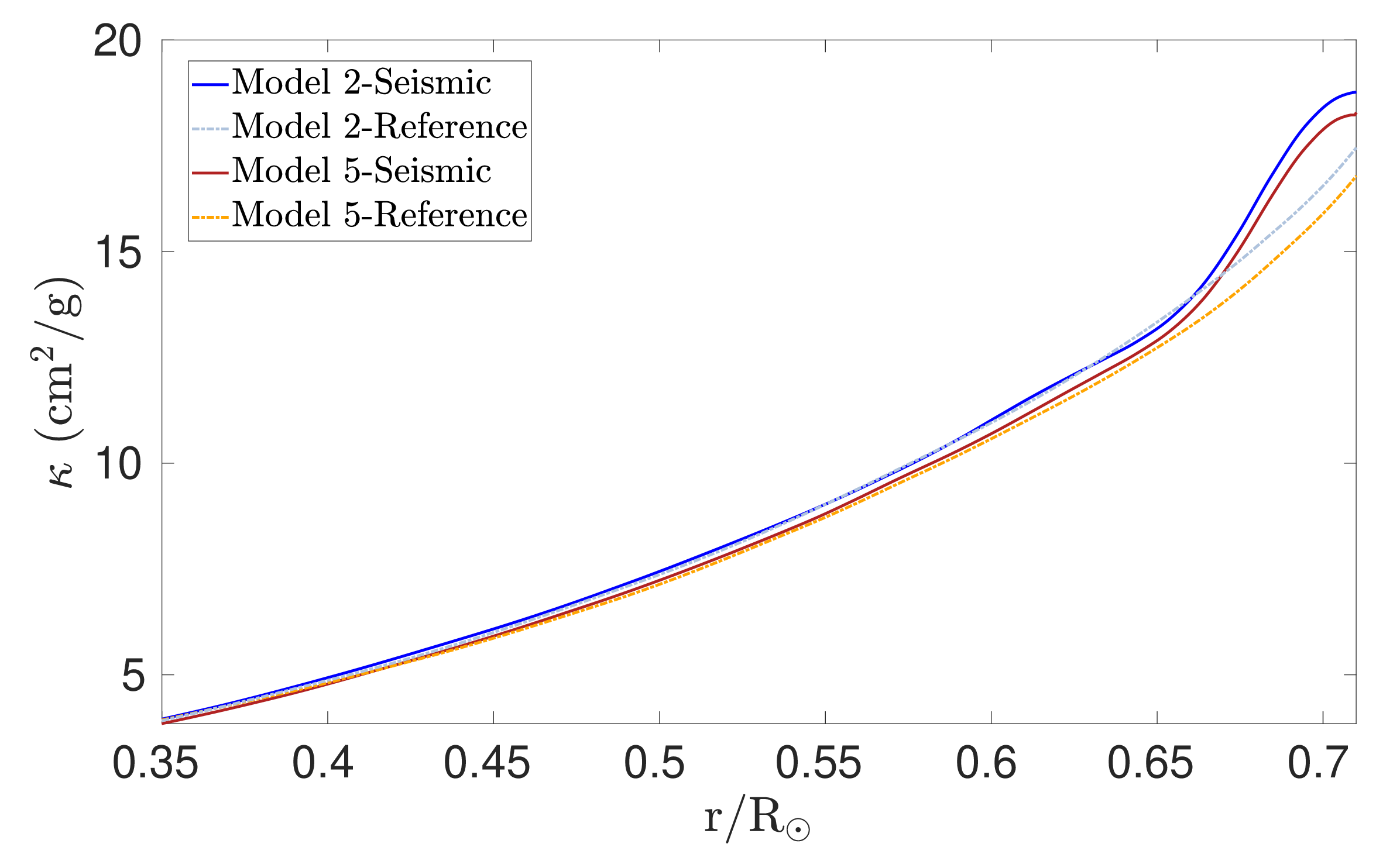}
	\caption{Effects of reference EOS and opacities on the seismic profiles. Blue curves are related to Model 2 using AAG21 abundances and the SAHA-S EOS, red and orange curves to Model 5, using the AGSS09 abundances and OPLIB opacities. Reference opacity profiles are indicated by dashed lines, while plain lines indicate the seismic opacity profiles.}
		\label{FigOpact}
\end{figure*} 

\begin{figure*}
	\centering
		\includegraphics[width=15cm]{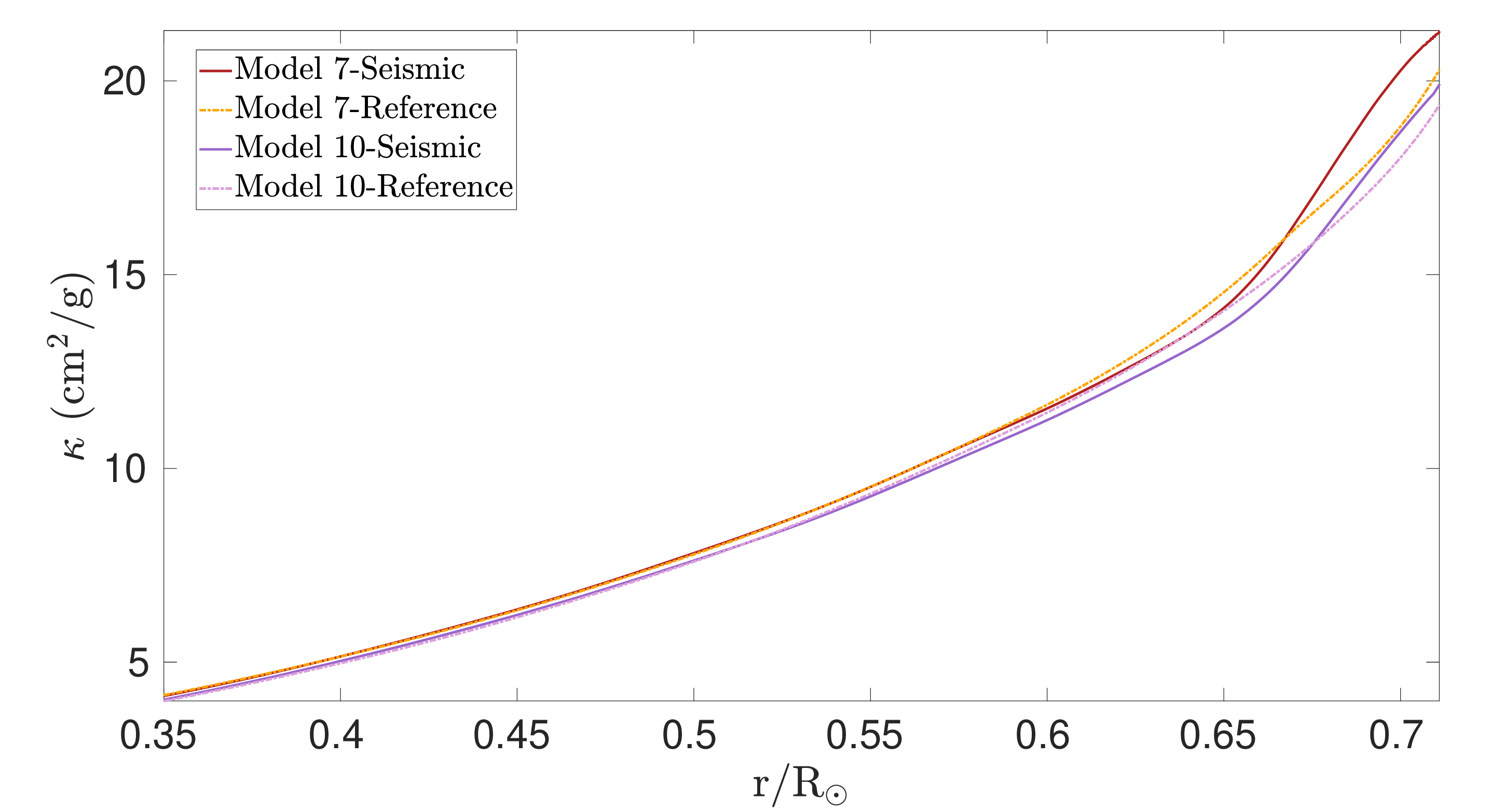}
	\caption{Effects of abundances on the seismic profiles. Purple curves are related to Model 10 using MB22 abundances, red and orange curves to Model 7, a standard solar model using GN93. Reference opacity profiles are indicated by dashed lines, while plain lines indicate the seismic opacity profiles.}
		\label{FigAbundance}
\end{figure*} 

Another crucial aspect of the reconstruction procedure is the assumption of a fixed equation of state for solar plasma. It comes into play when determining the temperature profile, T, from the given $\rho$, P, X, Z. The values of $\rho$ and P are obtained from the reconstruction procedure of Buldgen et al. (2020)\cite{Buldgen2020}, while X and Z are extracted from the evolutionary model. To assess the equation of state's impact on the final opacity determination, we use different tables available in CLES. As shown in Supplementary Fig. 7, the impact is minimal, mainly because the variations in the determined temperature in the solar interior's radiative layers are very small for various equations of state. This does not diminish the equation of state's overall importance in solar models and helioseismic inversions, as demonstrated in various studies \cite{Elliott1996,Basu1997,Elliott1998,Daeppen2006}. Rather, it highlights the current tables' high consistency, providing nearly identical temperatures for a given set of thermodynamic variables in the radiative layers. 

\begin{figure*}
	\centering
		\includegraphics[width=12cm]{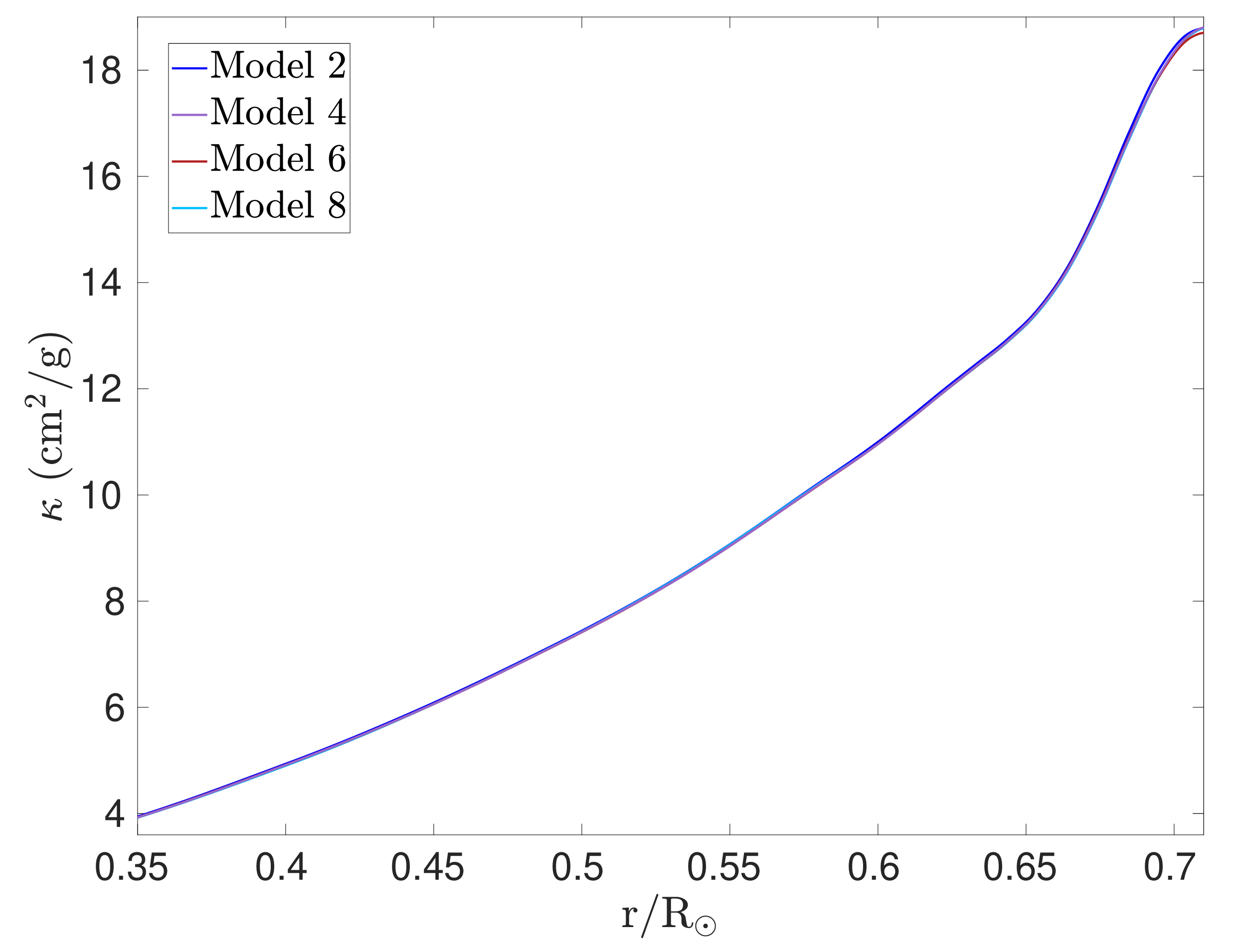}
	\caption{Effects of the equation of state, mixing parametrization and nuclear reaction rates on the seismic opacity profiles.}
		\label{FigEOS}
\end{figure*} 

Similarly, we test in Supplementary Fig. 8 the impact of using different opacity tables in the evolutionary models. We see that, despite starting from very different initial tables, the reconstruction procedure provides essentially the same seismic opacity profile, with again a discrepancy at the BCZ seen for every table. 

\begin{figure*}
	\centering
		\includegraphics[width=13cm]{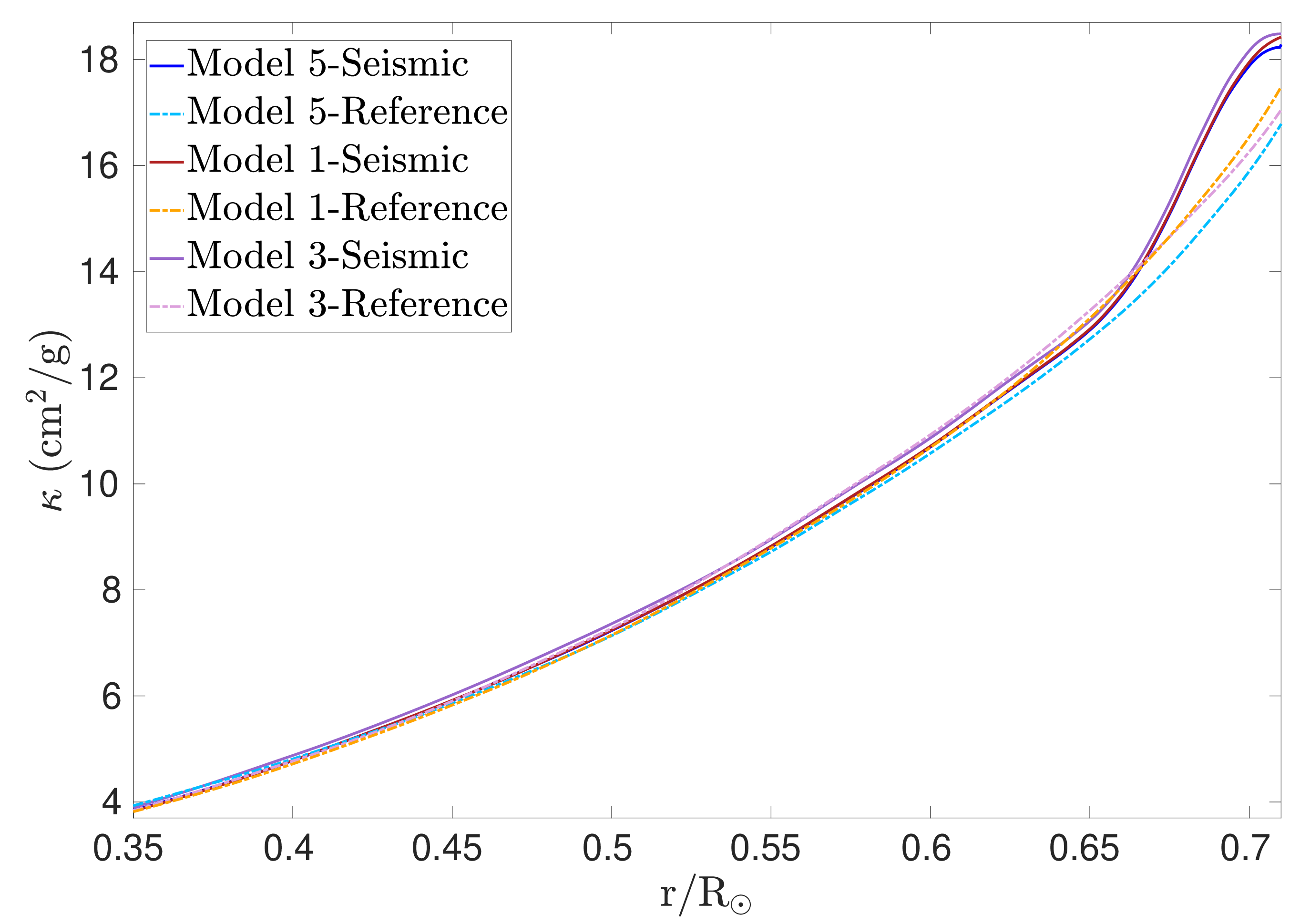}
	\caption{Effects of the reference opacity tables (Model 1: OPAS, Model 3: OP, Model 5: OPLIB) on the reference and seismic opacity profiles.}
		\label{FigEffectOP}
\end{figure*} 

The third assumption involves the nuclear reaction rates from our stellar evolution code. This, once again, has little to no impact on the seismic opacity profile, given that the most crucial parameter concerning energy production is the solar luminosity. To verify this, we use the NACRE rates\cite{NacreI} for Model 4, instead of the rates of Adelberger et al. (2011)\cite{Adelberger} used in all other models. Supplementary Fig. 7 demonstrates that the impact on the determined opacity is minimal.

While nuclear reaction rates do not impact the final results, we have the means to assess the properties of the core of the seismic models. Solar neutrino flux measurements provide insight into some limitations of our models. In practice, the properties of the solar core become decoupled from the opacity profile in the upper radiative layers, as long as the correct luminosity value is attained by the energy production. As a test case, we create a completely biased parameterization of the core's chemical composition, causing the model to be in full disagreement with measured neutrino fluxes and significantly differ in central chemical composition while still achieving the correct luminosity value. This model lacks physical meaning, but as it reproduces the correct luminosity value, Supplementary Fig. .9 illustrates that it still reaches the same opacity value as a more realistic model, demonstrating the robustness of our approach regarding neutrino flux measurements. However, this simple experiment also underscores the importance of neutrino flux measurements in resolving degeneracies in modeling the solar core (see e.g., \cite{Haxton2013,Gough2019,Villante2021} and references therein).

\begin{figure*}
	\centering
		\includegraphics[width=16cm]{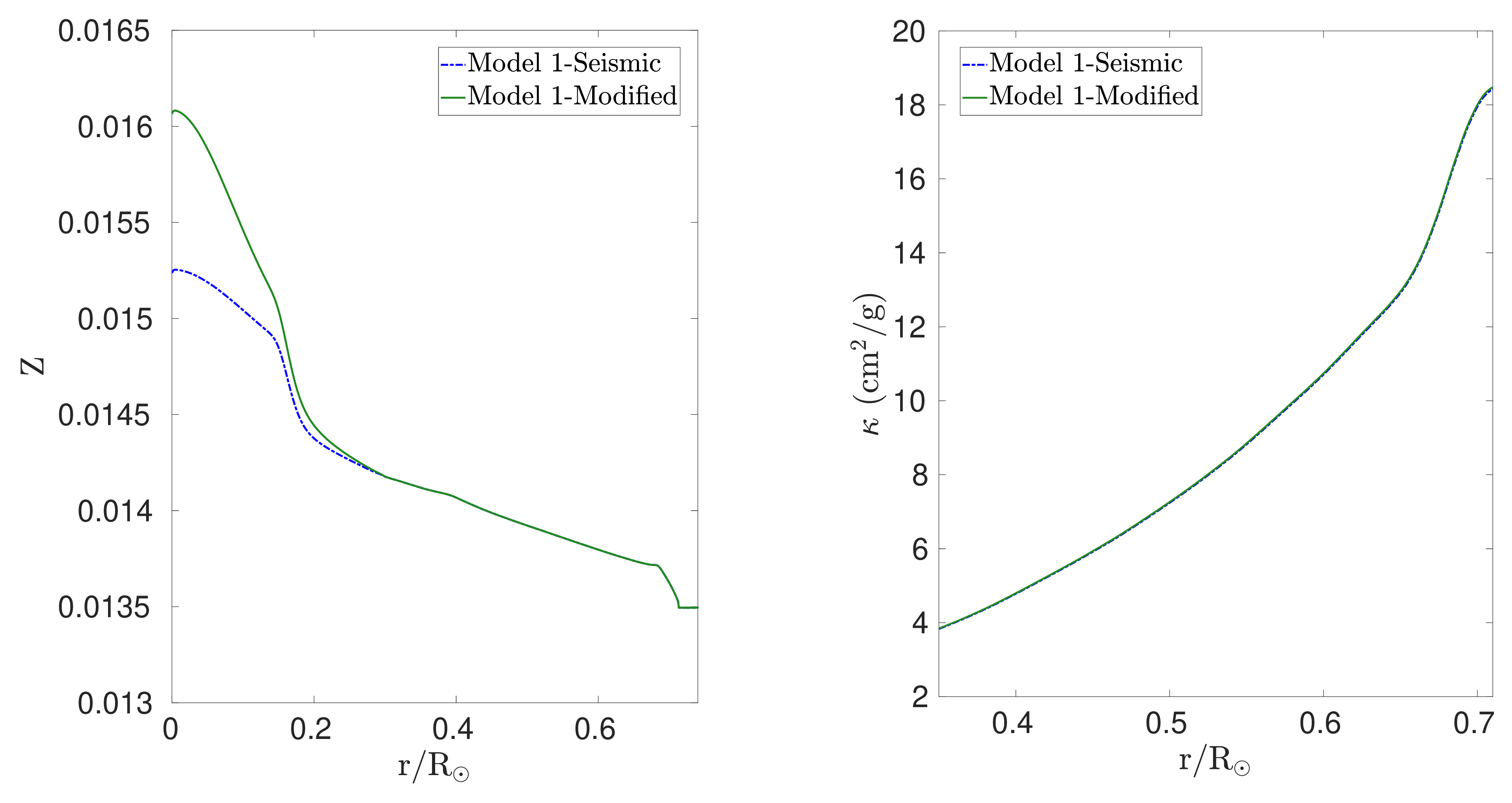}
	\caption{\textit{Left panel:} variations of the metallicity profile for one of the seismic models in the opacity reconstruction, both models fit the solar luminosity value. \textit{Right panel:} opacity profiles in the upper radiative layers obtained for both metallicity profiles of the left panel. The curves are undistinguishable}
		\label{FigZWeird}
\end{figure*} 

The last test we perform to fully determine the robustness of our technique is to use a different dataset of acoustic oscillations for the full seismic reconstruction procedure. In all reconstructions so far, we used the dataset from Buldgen et al. (2020)\cite{Buldgen2020}, namely a combination of Michelson Doppler Image (MDI) and Birmingham Solar Oscillation Network (BiSON) data\cite{Basu2009, Davies2014}. Another chosen dataset may alter the Ledoux discriminant inversion and thus the determined opacity profile. To check this impact, we use different MDI datasets\cite{Larson2015} and carry out a full seismic reconstruction. The final results are illustrated in Supplementary Fig. 10 for Model 2, showing that slight changes can be expected in the reconstructed opacity, but that the deviations close to the BCZ are still largely dominant and significant. 

\begin{figure*}
	\centering
		\includegraphics[width=12.5cm]{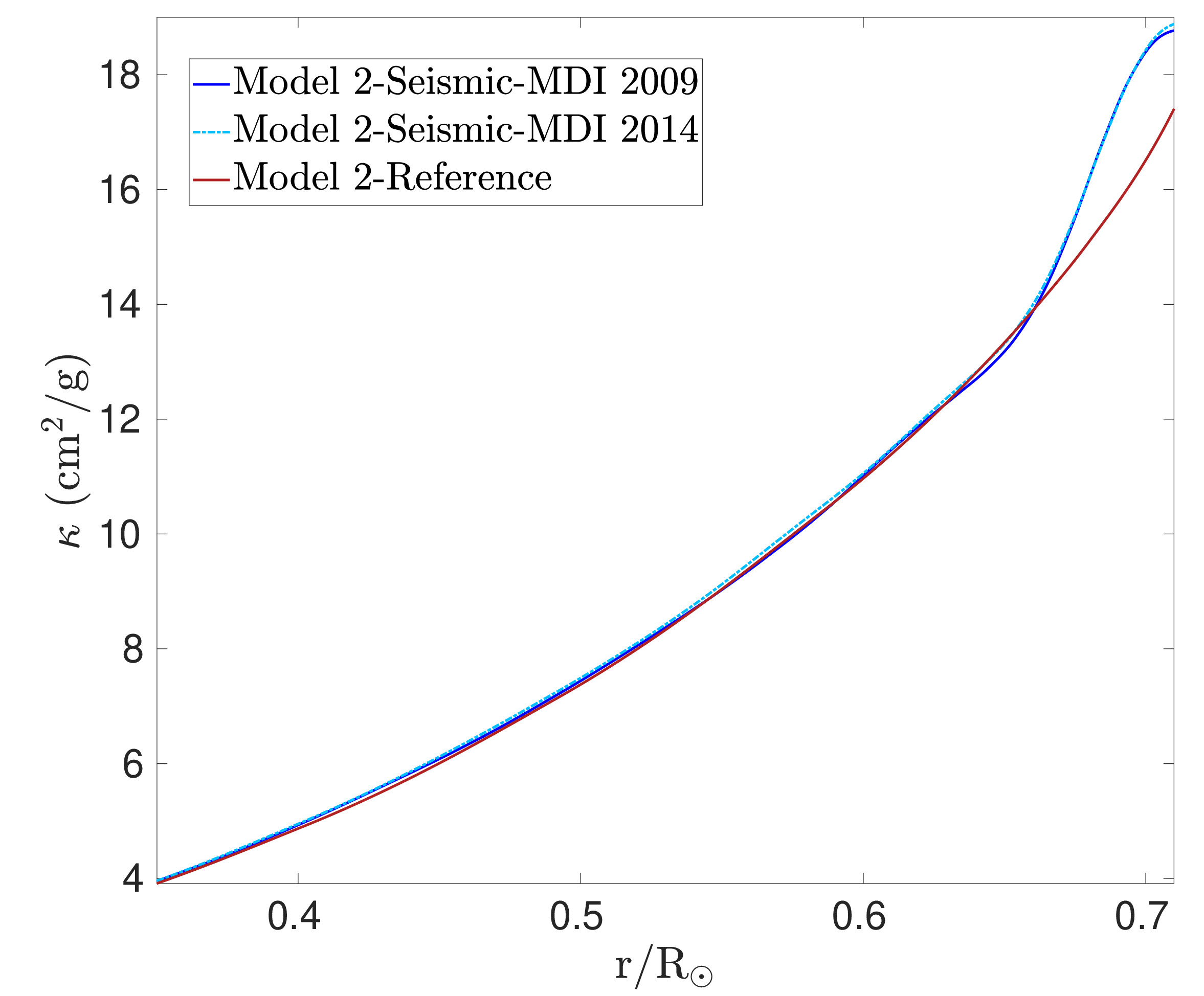}
	\caption{Changes in the reconstructed opacity profile for Model 2 using the 2009 Michelson Doppler Imager (MDI) data and the 2014 MDI data.}
		\label{FigMDIData}
\end{figure*} 

To summarize, the main assumption of the method is the underlying chemical composition profile of the model in the upper radiative layers. This profile may vary due to a change in the underlying physical properties of the model: individual relative abundances, transport of chemicals, reference opacity tables for the calibration procedure (although some models leading to a too low helium abundance in the CZ could be rejected on that base). The equation of state and nuclear reactions have almost no impact on the results, while the dataset used for the inversion has a limited impact ($<1\%$) on the inferred opacity. None of these assumptions alleviate the need for an opacity increase at the BCZ.

\subsection*{Theoretical opacity computations}
In this section, we first present in Supplementary Figure 10 and 11 in the Supplementary Information the contributions to the mean Rosseland opacity of less abundant elements of the solar mixture such as sulfur, or nickel but also of abundant elements such as hydrogen, helium or carbon which have lower contributions at the BCZ for both OPAS and SCO-RCG on the studied seismic thermodynamical path. We compare these computations to those based on version 3.3 of the OP code\cite{Badnell2005}, that is  cited as the reference opacity table of Standard Solar Models \cite{Vinyoles2017} and to the OPLIB opacities\cite{Colgan}. The comparisons are made for equal thermodynamical conditions of the models, namely Model 1 for OPAS and Model 2 for SCO-RCG.

Looking at Supplementary Fig. 11, we observe that systematically higher opacities are found with SCO-RCG with respect to both OP and OPLIB, with some elements such as sulfur or carbon showing very significant increases. For these ions, the treatment of the Stark effect can probably be invoked to explain the large differences between the codes. Heavier elements such as nickel show a behaviour very similar to iron, perhaps pointing at a similar origin of the discrepancies between the opacity computations. Significant differences are seen for hydrogen, but not for helium, which could point at differences in the equation of state and plasma effects\cite{Pradhan2024}. Small modifications for nickel have also been found at high temperature for B-type stars\cite{HuiBonHoa2021}, while large discrepancies are found at low temperatures. Differences seen for key elements such as nickel and iron (see Fig. 4), would have a significant impact on the oscillations of B-type stars. This triggered studies on the required opacity modifications to explain the observed modes \cite{DD2010,DD2013,Moravveji2016,DD2017}. We emphasize here that such opacity modifications in these stars can also result from the effects of radiative accelerations, in addition to the modifications of the opacities themselves\cite{Mombarg2022}.

\begin{figure*}
	\centering
		\includegraphics[width=15.5cm]{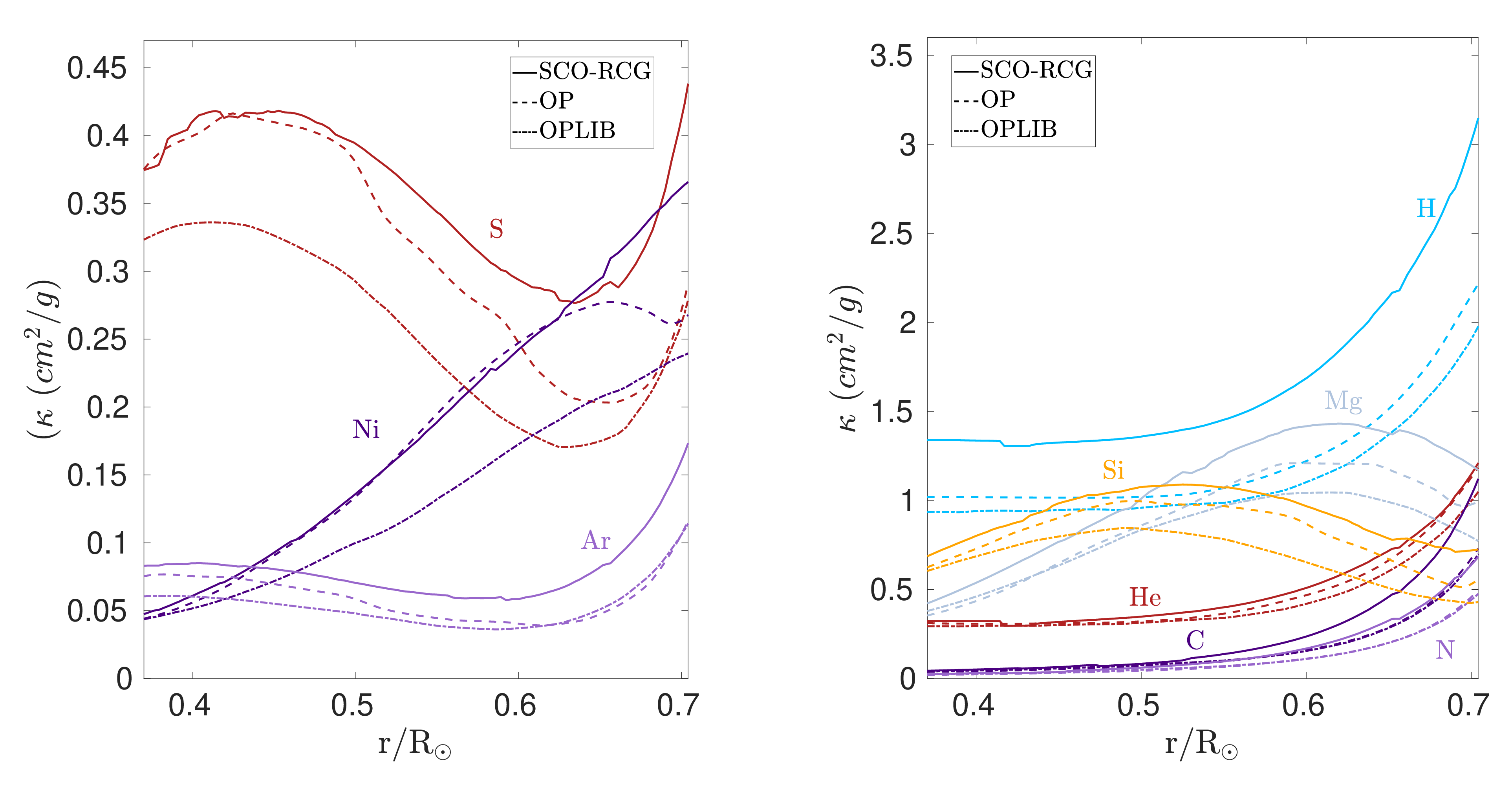}
	\caption{Opacity profiles for secondary contributors at the base of the solar convective zone on the seismic thermodynamical path for both the OP and SCO-RCG codes. \textit{Left panel:} contributions of sulfur, nickel and argon. \textit{Right panel:} contributions of hydrogen, magnesium, silicon, carbon, helium and nitrogen.}
		\label{FigOpacContribSuppSCORCG}
\end{figure*} 

Comparing OPAS, OP and OPLIB in Supplementary Fig. 12, we find that the differences are much smaller than in Supplementary Fig. 11. In general, OPAS opacities will be smaller than the OP ones. OPLIB and OPAS tend to agree for all lighter elements up to Nitrogen, but systematically lower opacities are found for heavier elements. The differences between OPAS and OP are particularly apparant in hydrogen and sulfur, while nickel shows some deviations in OP at the BCZ. This is not completelely unexpected as nickel was extrapolated from iron data in OP, but OPLIB and OPAS are not in good agreement for this element either, calling for further investigations. We observed the same pattern for hydrogen and helium, namely that helium seems to match very well between OP, OPLIB and OPAS, but hydrogen shows significant deviations throughout the structure, with OP having systematically higher values than both OPAS and OPLIB. While these differences remain limited, they still have an impact on the chemical composition of a calibrated model, as seen in Table 1 for Model 1 and Model 3 showing differences of $0.003$ in helium mass fraction, which is significant at the level of precision of helioseismic constraints (as $1$ $\sigma=0.0035$ \cite{BasuYSun}). 

\begin{figure*}
	\centering
		\includegraphics[width=15.5cm]{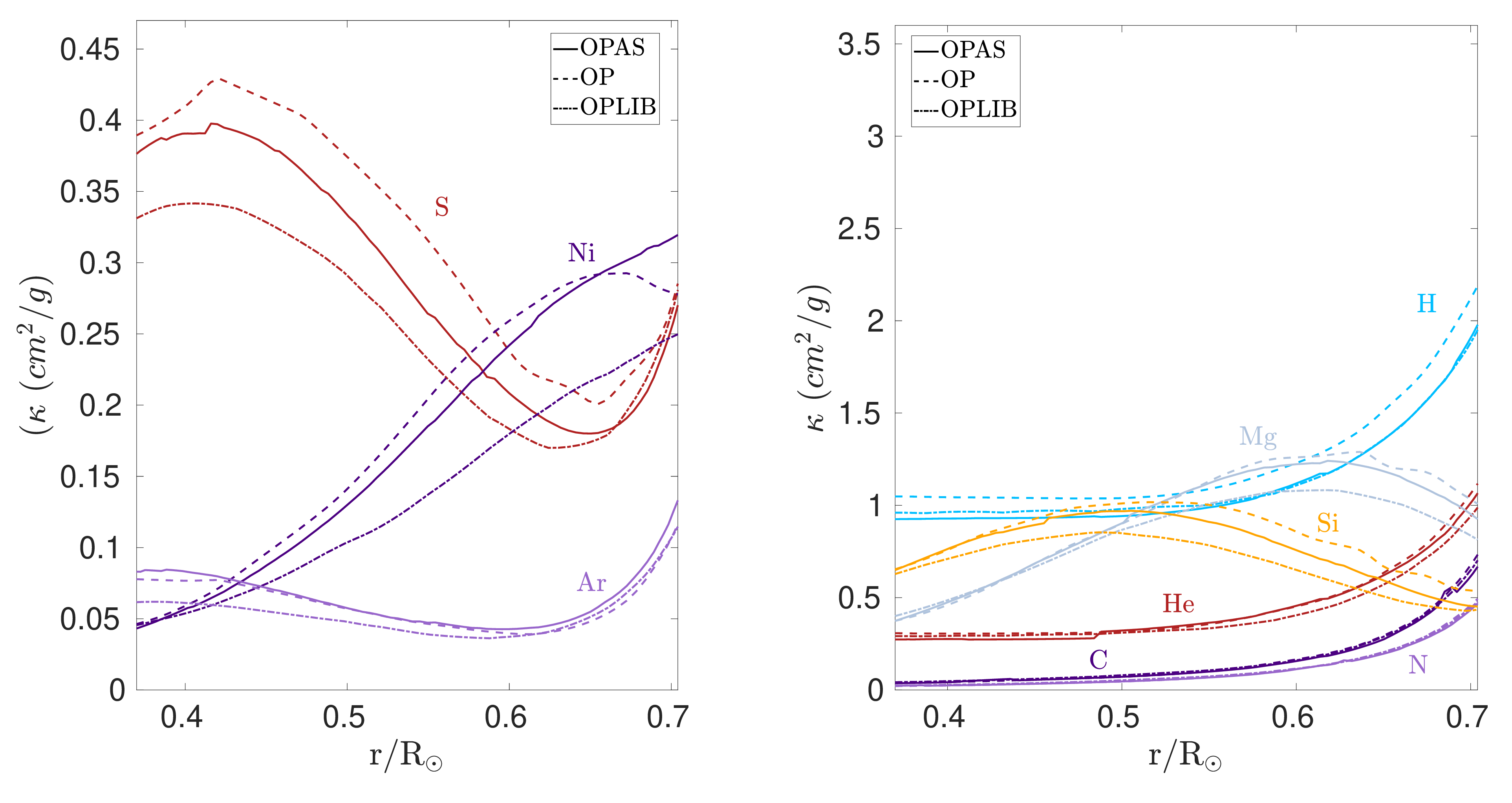}
	\caption{Opacity profiles for secondary contributors at the base of the solar convective zone on the seismic thermodynamical path for both the OP and OPAS codes. \textit{Left panel:} contributions of sulfur, nickel and argon. \textit{Right panel:} contributions of hydrogen, magnesium, silicon, carbon, helium and nitrogen.}
		\label{FigOpacContribSuppOPAS}
\end{figure*} 

\subsubsection*{SCO-RCG computations}
SCO-RCG \cite{Pain2015,Pain2021} is an opacity code combining fine-structure calculations with statistical modelling of radiative transitions in local-thermodynamic-equilibrium (LTE) plasmas. The code combines the data required for computing detailed transition arrays in the superconfiguration formalism (SCO: Superconfiguration Code for Opacity) within a plasma model accounting for the density and screening effects on the wavefunctions. The level energies and the line energies and strengths are computed by the module RCG (Robert Cowan's ``G'' subroutine), which is an ingredient of Cowan's atomic structure code \cite{Cowan1981}. RCG proceeds to the diagonalization of the Hamiltonian. 

\textbf{Accounting for levels, configurations and superconfigurations}\\
The computation is initialized an average-atom calculation, providing the mean populations of the subshells and a list of superconfigurations:

\begin{equation}
(1s)^{q_1}(2s)^{q_2}\cdots (n_{k-1}\ell_{k-1})^{q_{k-1}}\sigma^{q_k}
\end{equation}

\noindent is built, where
\begin{equation}
\sigma=\left(\prod_{i=k}^Nn_i\ell_i\right),
\end{equation}

\noindent where $n_N\ell_N$ is, at a given temperature and density, the highest-energy bound subshell predicted by the average-atom calculation. The LTE fluctuation theory is then applied to the average-atom non-integer mean populations to establish the range of variation for the populations $q_k$, where $k=1,N$. This process defines the possible configurations (if $q_k=0$) or superconfigurations (if $q_k>0$). The superconfigurations are subsequently sorted based on their respective Boltzmann probabilities, initially estimated using the average-atom wavefunctions and eigen-energies. Next, a self-consistent calculation is performed for each superconfiguration, which has therefore its own potential and corresponding set of wavefunctions and eigen-energies, used to reevaluate the probabilities.\\
The power of this hybrid approach lies in its ability to account for numerous highly excited states and satellite lines. Although their probabilities may be small, their large number means they are likely to make a significant contribution to the total opacity. In SCO-RCG, orbitals are handled individually up to a specified limit, beyond which they are merged into a single supershell (referred to as the Rydberg supershell). The grouped orbitals (in this case, those with $5<n<16$) are selected to be weakly interacting with inner orbitals. A Detailed Line Accounting (DLA) calculation is then performed for all transition arrays starting from that configuration. DLA calculations are only carried out for pairs of configurations that give rise to fewer than 800,000 lines (with the maximum size of a $J$-block within a configuration set to 4,000). For all other cases, transition arrays are represented statistically using Gaussian profiles within the frameworks of Unresolved Transition Arrays (UTA) \cite{Bauche1979} or Spin-Orbit Split Arrays (SOSA) \cite{Bauche1985}. If the supershell contains at least one electron, transitions originating from the superconfiguration are treated using the Super Transition Arrays (STA) model \cite{Barshalom1989}, ensuring that no configuration is omitted. The contribution of the supershell is thus kept as small as possible.
The computational effort in SCO-RCG is now dominated by these detailed calculations. As a result, the computed spectrum becomes less sensitive to the modeling of the remaining statistical contributions (UTA, SOSA, STA). The Partially Resolved-Transition-Array (PRTA) model \cite{Iglesias2012} has also been implemented, allowing for the replacement of many statistical transition arrays with smaller-scale DLA calculations. These DLA computations are carried out using the wavefunctions of the "real" configuration that was previously calculated. The electrostatic variance due to passive subshells is added to the widths of individual lines of the DLA calculation. This approach has been extended to the STA formalism by temporarily excluding the Rydberg supershell during the calculation, and adding its contribution to the widths of all lines as a Gaussian ``dressing function''. For instance, the calculation of the iron opacity in the solar mixture at the boundary of the solar convective zone involves about 1,000 non-relativistic ($n\ell$) configurations, 75,000 relativistic ($n\ell j$) configurations, and 2 billion J-levels. The total number of lines in the spectrum is approximately one billion, corresponding to 20,000 DLA transition arrays. The statistical part includes 460,000 UTAs, SOSAs, and STAs.

\textbf{Recent developments: inverse Bremsstrahlung and Stark effect}\\

The inverse Bremsstrahlung (free-free) absorption cross-section is computed following the approaches of Hummer \cite{Hummer1988} and van Hoof \cite{vanHoof2014}. The thermal averages are performed numerically, and the resulting values are then fitted by a two-dimensional Chebyshev expansion in temperature and photon energy, both in a logarithmic scale.
The included line-broadening mechanisms comprise inelastic collisions in the impact approximation, the Stark effect from neighboring ions in the quasi-static approximation, the Doppler effect, and the natural line width. The elastic contribution from electron collisions is neglected because it is much smaller than the inelastic one. The Doppler effect is modeled using a Gaussian profile, while inelastic collisions with electrons are described by a Lorentzian profile. Except for specific lines of one- and two-electron atoms (as discussed below), the ionic Stark effect is modeled by a Gaussian profile. The corresponding half-widths at half maximum (HWHM) are denoted by $\gamma_{\mathrm{Doppler}}$, $\gamma_{\mathrm{coll}}$, $\gamma_{\mathrm{Stark}}$ and $\gamma_{\mathrm{nat}}$, respectively. The convolution of the Gaussian and Lorentzian profiles leads to a Voigt profile:
\begin{equation}
\Psi(E)=\frac{1}{\sqrt{2\pi}\sigma}\mathcal{V}\left(\frac{E}{\sqrt{2}\sigma},\frac{a}{\sqrt{2}\sigma}\right),
\end{equation}
$a=\gamma_{\mathrm{coll}}+\gamma_{\mathrm{nat}}$ being the Lorentzian parameter, $E=h\nu$ the photon energy, $\sigma=(\gamma_{\mathrm{Stark}}+\gamma_{\mathrm{Doppler}})/\sqrt{2\ln 2}$ the standard deviation of the Gaussian, and $\mathcal{V}$ the Voigt function:
\begin{equation}
\mathcal{V}\left(x,y\right)=\frac{y}{\pi}\int_{-\infty}^{\infty}\frac{e^{-u^2}}{y^2+(u-x)^2}du,\label{voigt}
\end{equation}
computed, in SCO-RCG, following the numerical procedure described in Ref. \cite{AVRETT1969}.
\vspace{5mm}\\
\indent {\it Inelastic electron collisions}\\
The electron collisions are described following Dimitrijevi\'{c} and Konjevi\'{c} \cite{DIMITRIJEVIC1980,DIMITRIJEVIC1986,DIMITRIJEVIC1987}, which relies on Baranger's expression for the width of an isolated line \cite{BARANGER1958,BARANGER1962}:
\begin{equation}
\gamma_{\mathrm{coll}}=n_e\left\{v\left[\sum_{{\rm i'}}\sigma_{{\rm ii'}}+\sum_{{\rm f'}}\sigma_{{\rm ff'}}\right]\right\},
\end{equation} 
where $n_e$ is the density of free electrons, $v$ their velocity, and $\sigma_{{\rm ii'}}$ and $\sigma_{{\rm ff'}}$ are the inelastic scattering cross-sections for the initial and final states respectively. The sums run over all states interacting with the initial ${\rm i}$ and final state ${\rm f}$, and are weighted by the Boltzmann distribution for the free electrons. Only $\Delta n=0$ transitions are taken into account and the different contributions are weighted by an inelastic Gaunt factor which depends on the ratio between the free-electron thermal energy and the transition energy. Unlike \cite{DIMITRIJEVIC1980,DIMITRIJEVIC1986,DIMITRIJEVIC1987}, we do not use hydrogenic expressions, since the SCO-RCG code computes the radial integrals for each superconfiguration. The contribution of the elastic collisions to electron broadening is not included in SCO-RCG, as it is usually much weaker than that of inelastic collisions\cite{ROZSNYAI1977}.
\vspace{5mm}\\
{\it Ionic Stark effect}\\
The treatment of the ionic Stark effect implemented in the first version of SCO-RCG was proposed by Rozsnyai \cite{ROZSNYAI1977}. The broadenings of the initial and final states are assumed independent. 
{\it Specific ionic Stark model for Lyman ${\rm (Ly)}$ and Helium ${\rm (He)}$ lines}\\
A more accurate modeling of the Stark effect is used for hydrogen- and helium-like ions \cite{PAIN2017a,PAIN2017b}. The line profile then reads
\begin{equation}
\phi(\nu)\propto\frac{1}{\pi}\int\mathrm{Re}\left[\mathrm{Tr}\{\hat{d}.\hat{X}^{-1}\}\right]W(F)dF,
\end{equation}
where $\hat{X}=2i\pi\left(\nu+\nu_1\right)-i\hat{H}(F)/\hbar-\hat{\Lambda}_c$, $\nu_1$ is the frequency of the lower state and $\hat{H}(F)=\hat{H}_0-\hat{d}.F$ is the Hamiltonian of the ion subject to an electric field $F$, which follows the normalized distribution $W(F)$. The low-frequency microfield distribution is parameterized by the electron-ion screening constant and the ionic coupling parameter, based on simulations by Potekhin \emph{et al.} \cite{POTEKHIN2002}. The analytical formulas, which accurately reproduce the calculated electric microfield probability distributions, are expected to be valid for both neutral and charged plasma point particles with values of $\Gamma$ ranging from 0 to 100, and for a wide range of effective electron screening lengths. $\hat{H}_0$ represents the Hamiltonian without the electric field, while $\hat{d}$ and $\hat{\Lambda}_c$ denote the dipole and collision operators, respectively. The latter is derived from Griem \emph{et al.} \cite{GRIEM1959}, assuming classical straight-path electron trajectories that do not induce transitions in the radiator. The lower impact parameter is chosen to be the Bohr radius of the shell, and the upper one is the Debye length. The trace (Tr) operation is performed over all the states of the upper level. The line profile can be expressed as a sum of Voigt functions (see Eq. (\ref{voigt}) in Ref. \cite{HUMLICEK1979}).

\subsubsection*{OPAS computations}

The latest major developments of the OPAS code\cite{Blancard2012,Mondet} concern the line shape modelling and the description of the free-free component of the spectral opacity.

Fluctuating electric fields produced by the free electrons and the slow-moving ions in hot dense plasmas are known to have an important influence on the line profiles. In order to take the Stark effect into account on both the bound states and radiative data of partially ionized atoms and assuming that the perturbing electric field induced by the ions in the plasma can be described using the quasi-static approximation, we developed a model to calculate relativistic atomic structure in an external static electric field. This approach takes into account configuration interactions in a very general way. N-electron eigenfunctions are deduced from the diagonalization of the Dirac hamiltonian to which is added the interaction with the electric field. Eigenfunctions are expanded in terms of the Slater determinant using a Slater states expansion technique adapted from Eissner and Nussbaumer\cite{Eissner1969}. The one-electron wavefunctions are Dirac spinors resulting from the Dirac equation in an effective central potential\cite{SHADWICK198995}. Depending on the electric field, transition energies and line strengths are then calculated. The resulting spectrum is computed assuming that each line is dressed by a Voigt profile including Doppler, natural, and electron impact broadenings.  This model is applied, in the present work, for hydrogen-like and helium-like ions. For the other ionic stages, a statistical ionic Stark broadening is applied using a model proposed by Rozsnyai\cite{ROZSNYAI1977}. Following Iglesias and co-workers, a Gaussian cutoff is applied to the Voigt profile in order to remove its unphysical far wing behavior\cite{Iglesias2009}. The electric fields we considered are sampled from an ion microfield distribution deduced a from self-consistent approach for astrophysical and laboratory plasmas (SCAALP) calculations\cite{Blancard2004, Faussurier2010, Faussurier2003, Laulan2008}.

The free-free component of the spectral opacity is computed from the free-free component of the real part of the frequency-dependent electrical conductivity deduced from the Kubo-Greenwood approach\cite{Faussurier2015}. In the present work, a hydrogenic Gaunt factor\cite{Hummer1988} is used to account for quantum corrections at high frequency. Such a correction is here justified, as the plasmas are essentially non-degenerate along the considered thermodynamical paths.

\subsubsection*{OPLIB/ATOMIC computations}

The Los Alamos OPLIB opacity database has been publicly available for more than forty years, and is currently accessible at the website\footnote{\url{http://aphysics2.lanl.gov/opacity/lanl}}. The website can produce monochromatic, multigroup and gray opacities for either pure elements or arbitrary mixtures. The most recent database release\cite{Colgan}, which is the version considered in the present work, was generated with the ATOMIC code.
ATOMIC is a multi-purpose plasma modeling code \cite{Magee2004,Hakel2006,Fontes2015} that can be run in local-thermodynamic-equilibrium (LTE) or non-LTE mode to calculate the atomic-level populations. These populations are calculated using the occupation-probability formalism within the ChemEOS equation-of-state model to smoothly dissolve atomic states into the continuum \cite{Hakel2004,Hakel2006,Kilcrease2015}. The fundamental atomic data, such as wavefunctions and level energies, are calculated with the semi-relativistic Hartree-Fock method\cite{Abdallah1988,Cowan1981} in the Los Alamos suite of atomic physics codes\cite{Fontes2015}.

The resulting line (bound-bound) contributions to the opacities are calculated in fine-structure detail. Line broadening for H- and He-like ion stages includes the Stark treatment of Lee (1988)\cite{Lee1988}, while electron collisional broadening \cite{Armstrong1966} is used for ion stages with three or more bound electrons. The photoionization cross sections that are used in the bound-free contributions to the opacities are calculated using the distorted-wave approximation\cite{Clark1991}. Additional details about the latest OPLIB release are provided in Colgan et al. (2016)\cite{Colgan}. 

\section*{Data Availability Statement}
\small{OPLIB Data are available at: \url{https://aphysics2.lanl.gov/apps/astro_atomic.py}. The codes used to generate the data are not publicly available due to constraints imposed by LANL. The SCO-RCG data generated in this study are available upon request to J.-C. Pain (jean-christophe.pain@cea.fr). The OPAS data itself is available upon e-mail request to philippe.cosse@cea.fr. The source data for the figures in the main file as well as the supplementary are available with the paper.}\\
\small{The dataset used for the inversions is publicly available from the Birmingham Solar Oscillation Network website\footnote{\url{http://bison.ph.bham.ac.uk/portal/frequencies}, we used here both 2009 and updated 2014 datasets)} and the Joint Science Operations Center portal \footnote{\url{http://jsoc.stanford.edu/MDI/MDI_Global.html}}.}\\
 The frequency sets can be found on \footnote{\url{https://cdsarc.cds.unistra.fr/viz-bin/cat/J/A+A/681/A57}}. The datasets generated during and analysed during the current study are available from the corresponding author upon request. Opacity profiles of the reference models are provided in Supplementary Data - 1 to 10, with the number corresponding to the Model number in Table 1. Data for all figures are provided in the Source Data with this manuscript - Figure 1 to 4, including the Supplementary Figures 1 to 12.
\section*{Code Availability Statement}
\small{The CLES stellar evolution code is a proprietary software, detailed structures are available on demand by sending an email to the corresponding author. Further information on the code and availability can made upon specific request by sending an email to the corresponding author and the main developer of the code, R.Scuflaire@uliege.be. The inversion software used is publicly available at\\ \url{https://lesia.obspm.fr/perso/daniel-reese/spaceinn/index.html}. The codes used to generate the OPAS data are not publicly available due to constraints imposed by CEA. OP tables, code are publicly available and online computations can be carried out from their website\\ \url{https://cds.unistra.fr/topbase/TheOP.html}}.

\section*{Acknowledgments}
GB acknowledges fundings from the SNF AMBIZIONE grant No 185805 (Seismic inversions and modelling of transport processes in stars) and from the Fonds National de la Recherche Scientifique (FNRS) as a postdoctoral researcher. M.D. acknowledges support from the Centre National d'Etudes Spatiale (CNES), focused on the PLATO mission. AP acknowledges partial support from a grant by the US National Science Foundation. PE has received fundings from the European Research Council (ERC) under the European Union’s Horizon 2020 research and innovation programme (grant agreement No 833925, project STAREX). Funding for the Stellar Astrophysics Centre was provided by The Danish National Research Foundation (Grant DNRF106). JC, CJF, PH and DPK were supported by the U.S. Department of Energy through the Los Alamos National Laboratory, which is operated by Triad National Security, LLC, for the National Nuclear Security Administration of U.S. Department of Energy (Contract No. 89233218CNA000001). We acknowledge support by the ISSI team ``Probing the core of the Sun and the stars'' (ID 423) led by Thierry Appourchaux. 

\section*{Author Contributions}
G.B. led the project, developed the inversion technique and carried out the analysis. J.C.P. and F.G. provided the SCO-RCG data and the associated description of the software. P.C. and C.B. provided the OPAS data and provided the associated description of the software. A.P. provided the OP data and provided the associated discussions of the various element contributions as well as discussion of the seismic opacity. C.J.F., J.C., P.H. and D.P.K provided the OPLIB data and the associated description of the software. A.N. and R.S. provided help with improvements of the Liège Stellar Evolution Code (CLES). J.C.D. provided help with the interpretation of the helioseismic results and the context of the study. M.D. and Y.L. verified the claims regarding radiative accelerations with the CESAM2k20 evolution code. S.V.A., V.A.B. and A.V.O provided the SAHA-S equations of state tables used in CLES and helped interpret the helioseismic results. C.P. provided help regarding the transport by internal gravity waves and the related issues with the determination of chemical composition profiles. T.C. provided help with the interpretation of the helioseismic results. P.E. provided help with the transport by the magnetic Tayler instability. All authors have contributed to the interpretation of the results and in providing comments on the papers.  

\section*{Competing Interest}
The authors declare no competing interests.

\bibliographystyle{essai}
\renewcommand{\refname}{References}

\end{document}